\pdfoutput=1
\documentclass[hidelinks]{aa}
\usepackage{newtxtext}
\usepackage[varg]{newtxmath}
\usepackage{booktabs}
\usepackage{scrextend}
\usepackage{orcidlink}
\usepackage{natbib,twoopt}
\usepackage{hyperref}
\hypersetup{
    colorlinks,
    linkcolor={red!50!black},
    citecolor={blue!50!black},
    urlcolor={blue!80!black},
    pdftitle={The impact of the presence of water ice on the analysis of debris disk observations},
    pdfauthor={T. A. Stuber and S. Wolf}
}
\bibpunct{(}{)}{;}{a}{}{,} 

\begin{document}


\title{The impact of the presence of water ice on the analysis of debris disk
       observations}
\titlerunning{Impact of water ice on the analysis of debris disk observations}
\author{T. A. Stuber \orcidlink{0000-0003-2185-0525}
        \and S. Wolf \orcidlink{0000-0001-7841-3452}}
\authorrunning{T. A. Stuber and S. Wolf}
\institute{Institut für Theoretische Physik und Astrophysik,
           Christian-Albrechts-Universität zu Kiel,
           Leibnizstr. 15, 24118 Kiel, Germany\\
           \email{tstuber@astrophysik.uni-kiel.de}}

\date{Received 28 May 2021/
Accepted 14 November 2021}

\abstract{
 The analysis of debris disk observations is often based on the assumption of a
 dust phase composed of compact spherical grains consisting of astronomical
 silicate. Instead, observations indicate the existence of water ice in debris
 disks.}
{
 We quantify the impact of water ice as a potential grain constituent in debris
 disks on the disk parameter values estimated from photometric and spatially
 resolved observations in the mid- and far-infrared.}
{
 We simulated photometric measurements and radial profiles of debris disks
 containing water ice and analyzed them by applying a disk model purely
 consisting of astronomical silicate. Subsequently, we quantified the deviations
 between the derived and the true parameter values. As stars in central
 positions we discuss a \mbox{$\beta$ Pic} sibling and main-sequence stars with
 spectral types ranging from A0 to K5. To simulate observable quantities we
 employed selected observational scenarios regarding the choice of wavelengths
 and instrument characteristics.}
{
 For the \mbox{$\beta$ Pic} stellar model and ice fractions
 \mbox{$\geq 50\ \%$} the derived inner disk radius is biased by ice
 sublimation toward higher values. However, the derived slope of the radial
 density profile is mostly unaffected. Along with an increasing ice fraction,
 the slope of the grain size distribution is overestimated by up to a median
 factor of \mbox{$\sim 1.2$} for an ice fraction of \mbox{$90\ \%$}. At the
 same time, the total disk mass is underestimated by a factor of
 \mbox{$\sim 0.4$}. The reliability of the derived minimum grain size strongly
 depends on the spectral type of the central star. For an A0-type star the
 minimum grain size can be underestimated by a factor of \mbox{$\sim 0.2$},
 while for solar-like stars it is overestimated by up to a factor of
 \mbox{$\sim 4$ -- $5$}. Neglecting radial profile measurements and using
 solely photometric measurements, the factor of overestimation increases for
 solar-like stars up to \mbox{$\sim 7$ -- $14$}.}
{}

\keywords{circumstellar matter -- interplanetary medium --
          infrared: planetary systems -- submillimeter: planetary systems --
          methods: numerical}

\maketitle


\section{Introduction}\label{introduction}

 Debris disks are optically thin, gas-poor circumstellar distributions of
 dust around main-sequence stars and brown and white dwarfs; the dust is
 continuously produced in collisional cascades initiated by colliding parent
 planetesimals \citep[for recent reviews of debris disks see][]
 {10.2458/azu_uapress_9780816531240-ch023,
 10.1146/annurev-astro-081817-052035, 10.1016/B978-0-12-816490-7.00016-3}.
 While debris disks were originally discovered in unresolved images by their
 infrared excess \citep{10.1086/184214, 10.1086/131620},
 today spatially resolved images of those systems are available.
 Paired with spectral energy distributions (SEDs) they allow
 us to break degeneracies in the analysis of debris disk observations and to
 constrain both the geometrical structure and properties of the dust
 distribution.

 At mid-infrared to millimeter wavelengths, where we trace the thermal emission
 of the dust, recent advancements in the field have become possible, thanks
 particularly to the \textit{Spitzer Space Telescope}
 \citep[\textit{Spitzer};][]{10.1086/422992}, the
 \textit{Herschel Space Observatory}
 \citep[\textit{Herschel};][]{10.1051/0004-6361/201014759}, and the
 \textit{Atacama Large Millimeter/submillimeter Array}
 \citep[ALMA;][]{2002Msngr.107....7K}. At shorter wavelengths, where we trace
 the stellar light scattered by the dust, using high-contrast imaging
 instruments such as the \textit{Hubble Space Telescope} (HST), the
 \textit{Gemini Planet Imager} \citep[GPI;][]{10.1117/12.672430}, the
 \textit{Subaru Coronagraphic Extreme Adaptive Optics}
 \citep[SCExAO;][]{10.1117/12.857818, 10.1086/682989}, and the
 \textit{Spectro-Polarimetric High-contrast imager for Exoplanets REsearch}
 \citep[SPHERE;][]{10.1051/0004-6361/201935251} helped to extend our knowledge.

 Among other parameters such as dust grain porosity
 \citep{10.1051/0004-6361/201220486, 10.1093/mnras/stw2675}, the chemical
 composition of the dust grains defines their absorption and scattering
 characteristics, and thus has a major impact on the appearance of a debris
 disk. However, to infer the dust composition from observations is difficult
 for various reasons. For example, characteristic dust features are found in
 the infrared, such as the prominent \mbox{$10\ \mu$m} or the
 \mbox{$69\ \mu$m} feature \citep[e.g.,][]{10.1038/nature11469} of silicate.
 However, debris disks frequently possess an inner cavity with a typical size of
 several astronomical units
 \citep[e.g.,][and references therein]{10.1086/306184, arXiv:astro-ph/0703383},
 and thus only small amounts of warm dust could produce these spectral
 features. Furthermore, the systems are often faint sources, which makes
 high-resolution spectroscopy of the scattered light and thermal emission
 radiation difficult. Apart from the silicate components, which are often
 modeled all together by the artificial material astronomical silicate
 \citep[hereafter silicate;][]{10.1086/162480, 10.1086/379123}, collisional
 debris can be composed of various other chemical species, including water ice
 \citep{10.1088/0004-637X/747/2/93}. Water ice is expected to be present in
 debris disks for several reasons. In protoplanetary disks, which are the
 direct progenitors of debris disks, water ice has been confirmed
 observationally
 \citep[][and references therein]{10.2458/azu_uapress_9780816531240-ch016}, and
 we know that planetesimals in our own Solar System can contain substantial
 amounts of water ice (ibid.). Direct observations of debris disks also suggest
 the presence of water ice. Around \mbox{HD 181327} \citet{10.1086/592567}
 found a broad emission peak at $60$ -- \mbox{$70\ \mu$m}, potentially caused
 by crystalline water ice. The analyses of several further debris disk
 observations favor a water ice--silicate mixture
 \citep[e.g.,][]{10.1051/0004-6361/201117731, 10.1051/0004-6361/201117714,
                  10.1088/0004-637X/776/2/111, 10.3847/0004-637X/831/1/97}.
 However, there is currently no solid evidence for water ice in a debris disk.
 A possible explanation is given by the effect of photosputtering by
 ultraviolet (UV) photons \citep{10.1051/0004-6361:20077686}, which might
 remove water ice from the grain surfaces within a short timescale. However,
 as we do not have secured evidence of this effect, we do not consider it
 in this study. If water ice is present in debris disks, its prominent
 \mbox{$\sim 3\ \mu$m spectral feature} will potentially be observable with
 the \textit{James Webb Space Telescope}
 \citep[JWST;][]{10.1007/s11214-006-8315-7, 10.1051/0004-6361/201936014}.

 In this study we evaluate the impact of water ice on the analysis of
 typical debris disk observations. Our approach is to simulate synthetic
 observations of ice-containing debris disks with a setup strongly motivated
 by real observations: photometric measurements and radial profiles extracted
 from spatially resolved images at wavelengths ranging from the mid-infrared to
 the millimeter regime. Our dust model employs ice--silicate mixture grains;
 the ice can undergo sublimation, which produces vacuum inclusions making the
 grain porous. Subsequently, we analyze these synthetic observations assuming
 compact silicate dust grains and compare the derived parameter values with the
 known true values (Sect.~\ref{sect_methods}). Debris disk parameters are then
 considered: the inner disk radius, the slope of the radial density and grain
 size distribution, the minimum grain size, and the total disk mass within the
 considered grain size interval. We present the results for dust
 configurations around a star comparable to $\beta$ Pic
 (Sect.~\ref{subsect_beta_pic_results}) and investigate the impact of the
 spatial resolution by varying the distance to the debris disk system
 (Sect.~\ref{subsect_dist_variation}). Subsequently, we
 explore the impact of the spectral type (A0 to K5,
 Sect.~\ref{subsect_spectral_types}). Finally, we summarize our results and
 relate them to observational analyses from the literature
 (Sect.~\ref{sect_discussion}).


\section{Methods}\label{sect_methods}


\subsection{General procedure}

 To address the general question outlined in Sect.~\ref{introduction}, our goal
 is to quantify the bias in the quantitative estimation of values of selected
 debris disk parameters if the presence of water ice (hereafter ice) is
 ignored. For this purpose we chose a procedure similar to that applied
 by \citet{10.1093/mnras/stw2675} to study the influence of dust grain porosity
 on the analysis of debris disk observations:
 \renewcommand{\labelenumi}{(\roman{enumi})}
 \begin{enumerate}
  \item Simulation of ice-containing debris disks; computation of SEDs and
        spatially resolved images;
  \item Simulation of synthetic observations
        \renewcommand{\labelitemi}{$\bullet$}
        \begin{itemize}
         \item SEDs: addition of artificial noise
         \item Images: convolution with an instrument beam, extraction of
         radial profiles, addition of artificial noise;
        \end{itemize}
  \item Simulation of reference debris disks consisting of pure compact
        silicate grains, computation of SEDs, extraction of radial profiles;
  \item Fitting of the synthetic observations (ii) using the reference
        observations (iii);
  \item Comparison of the best-fit parameter values (iv) with the correct ones
        (i).
 \end{enumerate}
 In the following the underlying debris disk model as well as the modeling and
 fitting approach are outlined.


\subsection{Grain properties and water ice model}\label{grain_model}

 Ice appears in multiple crystalline and amorphous phases depending on the
 environmental conditions. All phases
 are characterized by different physical characteristics, such as their optical
 properties. For example, the ice condensed onto grains in a cold interstellar
 or solar nebular environment is supposed to be amorphous. However, embedded in
 planetesimals and exposed to frictional and/or radioactive heating it can
 transform into a crystalline phase \citep[see][for a thorough review about
 ice physics]{physics_of_ice}.
 In our study we applied crystalline water ice with a density of
 \mbox{$\rho_{\mathrm{ice}} = 1.0$ $\mathrm{g \,cm}^{-3}$}
 \citep[e.g.,][]{10.1016/j.icarus.2008.02.005, 10.5047/eps.2009.03.001} at a
 temperature of \mbox{$55$ K} as a representative temperature in a cold debris
 disk.

 We employed a mixture of ice and silicate (hereafter ice-mixture), for
 which we used the data sets of wavelength-dependent complex refractive indices
 compiled by \citeauthor{10.1051/0004-6361/201936014}
 (\citeyear{10.1051/0004-6361/201936014}; see their Sect.~2 for a detailed
 description of the applied approach to merge the individual data
 sets), containing
 optical ice data of different wavelength regimes using the results of
 \citet{1998A&A...331..291L},
 \citet{10.3847/1538-4357/aac6d3},
 \citet{10.1364/AO.44.004102},
 \citet{10.1051/0004-6361/201424276},
 \citet{10.1093/mnras/sty2664}, and
 \citet{10.1364/AO.23.001206}.
 In their study, \citet{10.1051/0004-6361/201936014} obtained optical
 properties of nine ice-mixtures with an ice volume fraction $\mathcal{F}$
 ranging from 0.1 to 0.9 in steps of 0.1 by using the Maxwell--Garnett rule of
 effective medium theory \citep[EMT;][]{10.1098/rsta.1904.0024} with ice as the
 inclusion and silicate as the matrix. As complex refractive indices of silicate
 they used the data from \citet{10.1086/379123}. Likewise, we used this silicate
 data for the reference disks consisting of compact silicate grains. In this
 study we employed \mbox{$\rho_{\mathrm{sil}} = 3.8$ $\mathrm{g \,cm}^{-3}$} as
 the bulk density of silicate. Furthermore, we assumed all grains to be spherical
 and used the Mie \citep{10.1002/andp.19083300302} scattering tool \texttt{miex}
 \citep{10.1016/j.cpc.2004.06.070} to compute the required optical properties,
 such as their wavelength-dependent cross sections.

 Once released from a large parent planetesimal, the dust grains are exposed to
 a vacuum and stellar radiation, which may deplete the ice and modify its
 radial distribution. When exposed to stellar radiation, a grain heats up to an
 equilibrium temperature. Following \citet{10.1051/0004-6361/201936014}, we
 assumed a sublimation temperature of the ice inclusions of \mbox{$105$ K}. If
 the equilibrium temperature of an ice-containing grain exceeded the
 sublimation temperature, we assumed the leftover to be a porous silicate
 grain; that is, the ice is assumed to be replaced by a vacuum. The porosity
 $\mathcal{P}$, defined as the volume fraction of a vacuum, equals the ice
 fraction $\mathcal{F}$ of the precursor grain. The remaining porous silicate
 grain then has different optical properties, and thus different absorption and
 emission characteristics. Therefore, it settles to a different equilibrium
 temperature than its precursor grain. We derived the complex refractive
 indices of porous silicate as mixtures of silicate and a vacuum applying the
 EMT mixing rule.

 Highly energetic UV photons have the potential to deplete ice directly via
 photosputtering (or photodesorption).
 \citet{10.1051/0004-6361:20077686} find that around \mbox{$\beta$ Pic} only
 grains \mbox{$\gtrsim 20$ $\mu$m} can be at least partially icy at a distance
 \mbox{$\geq 40$ AU} from the star. However, the authors note that their
 results should be treated with care as they have not used full-scale
 collisional evolution simulations. Implementing their findings in our model
 might underestimate the importance of ice in young and collisionally active
 debris disks. Therefore, we decided to neglect UV photosputtering in this
 study. Thus, the results of this study are primarily applicable to
 collisionally active debris disks. The other extreme case, in which grains
 cannot retain their ice in the presence of UV photosputtering and only porous
 grains remain, has been considered by \citet{10.1093/mnras/stw2675}.
 \begin{figure}
  \resizebox{\hsize}{!}{\includegraphics{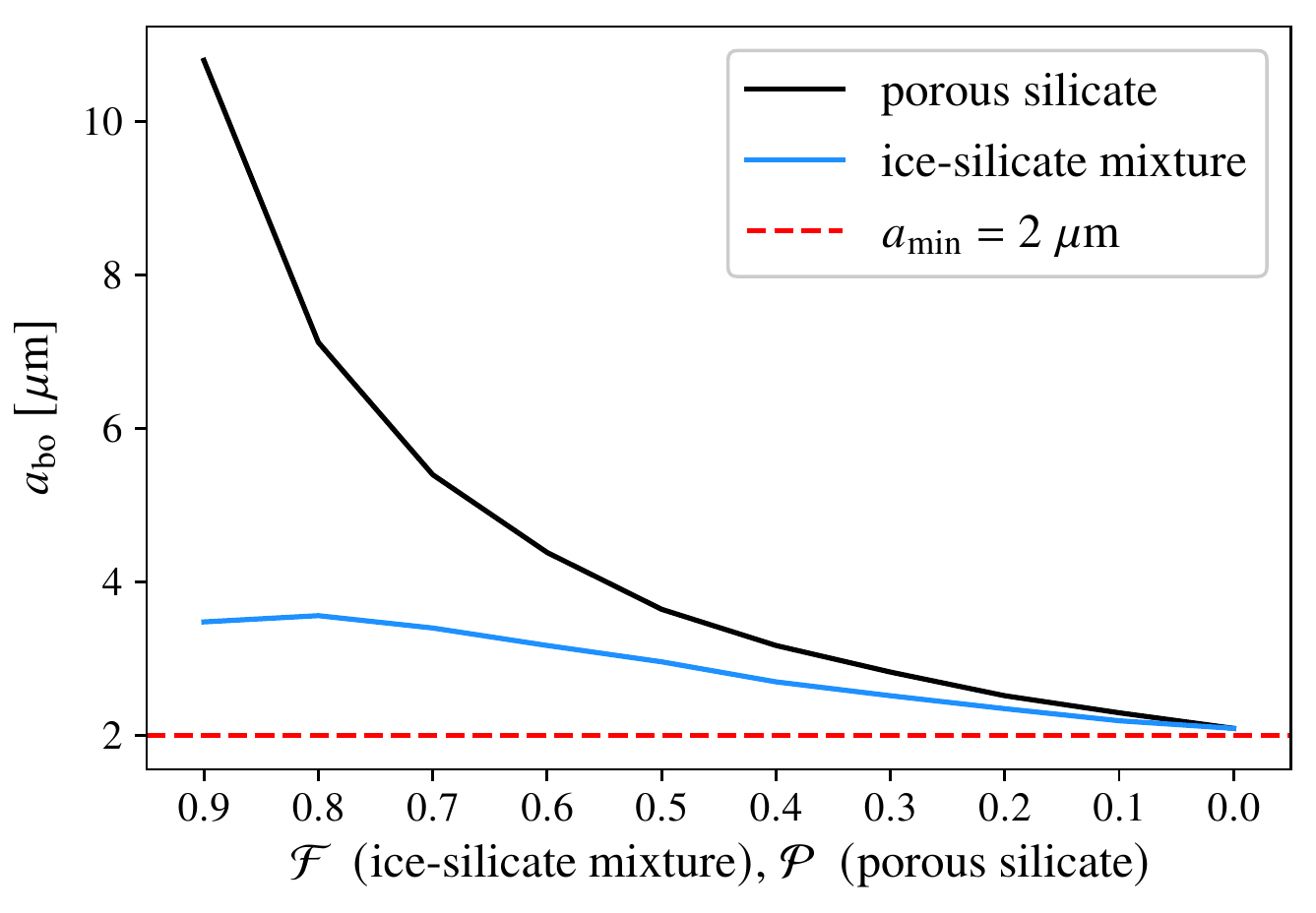}}
  \caption{Blow-out grain sizes $a_\textrm{bo}$ of grains consisting
           of an ice-mixture or porous silicate for different volume fractions
           of ice $\mathcal{F}$ in the case of ice-mixtures and a
           vacuum $\mathcal{P}$ in the case of porous silicate, respectively
           ($\mathcal{F} = \mathcal{P} = 0$: compact silicate). The central
           star is chosen comparable to $\beta$ Pic (see
           Sect.~\ref{stellar_model}). The dashed horizontal red line
           represents the minimum grain size chosen in the simulations,
           smaller than any possible blow-out grain size $a_\textrm{bo}$
           within our model space: \mbox{$a_\textrm{min} = 2$ $\mu$m}.}
  \label{fig_bo}
 \end{figure}

 The relative strength of the radiation pressure of the incident radiation onto
 a dust grain can be described by the $\beta$-factor, defined as the ratio of
 radiation to gravitational force
 \citep{10.1016/0032-0633(75)90078-1, 10.1016/0019-1035(79)90050-2}.
 We compute $\beta$ by taking the wavelength-dependence of the radiation
 pressure efficiency into account \citep{10.1051/0004-6361/201220486}
 \begin{equation}
  \beta \coloneqq \frac{F_{\mathrm{rad}}}{F_{\mathrm{gra}}}
        = \frac{3 \pi}{4 G c} \frac{R_{\star}^2}{M_\star}\frac{1}{a \rho}
          \int Q_{\mathrm{pr}} J_{\lambda}\left( T_{\star} \right)
               \mathrm{d} \lambda \; ,
 \end{equation}
 with the gravitational constant $G$, the speed of light in a vacuum $c$, the
 stellar radius $R_{\star}$, the stellar mass $M_{\star}$, the stellar effective
 temperature $T_{\star}$, the grain radius $a$, the bulk density $\rho$, the
 wavelength $\lambda$, the stellar spectrum $J_{\lambda}$
 (see Sect.~\ref{stellar_model}), and the wavelength-dependent radiation
 pressure efficiency factor $Q_{\mathrm{pr}}$.

 Assuming that the grains are released on a circular orbit, those with
 $\beta \geq 0.5$ become unbound, leave the system on parabolic or
 hyperbolic orbits
 \citep[see, e.g.,][]{10.1016/0032-0633(75)90078-1, 10.1051/0004-6361:20064907},
 and are removed from the simulations. The specific value of the blow-out grain
 size $a_{\textrm{bo}}$, below which all grains get blown out, is individual
 for each of the two materials of ice-mixture and porous silicate, denoted by
 $a_{\textrm{bo}}^\mathcal{F}$ and $a_{\textrm{bo}}^\mathcal{P}$, respectively.
 We assume the ice-mixture grains to reach thermal equilibrium before they get
 blown out of the system. Thus, if the grain temperature is above the ice
 sublimation temperature, first the ice is considered sublimated and second
 $\beta$ is calculated for the porous leftover. For dust around a star similar
 to \mbox{$\beta$ Pic} (see Sect.~\ref{stellar_model}) the blow-out grain sizes
 for both the ice-mixture and porous silicate grains are displayed in
 Fig.~\ref{fig_bo}.

 The grain size distribution is modeled as a power law
 $n\left( a \right) \propto a^{-\gamma}$, ranging from the minimum to maximum
 grain size, $a_{\textrm{min}}$ to $a_{\textrm{max}}$, with
 $n \left( a \right) \mathrm{d} a$ being the number of particles in the interval
 $\left[ a, a + \mathrm{d} a \right]$. A steady state collisional cascade would
 produce a size distribution with $\gamma = 3.5$ \citep{10.1029/JB074i010p02531}.
 Distinguishing between objects in the strength regime (diameter
 \mbox{$D \lesssim 0.1$ km}) and objects in the gravity regime
 (\mbox{$D \gtrsim 0.1$ km}) leads to values of $\gamma \sim 3$ -- $4$
 \citep{10.2458/azu_uapress_9780816531240-ch023}. We adopted this parameter range
 and explored it in this study. By using a simple power law for the grain size
 distribution, we neglect its wavy pattern caused by the blow-out of small
 grains by radiation pressure
 \citep{10.1016/0032-0633(94)90008-6, 10.1051/0004-6361:20077709,
        10.1007/s10569-011-9345-3}
 and effects caused by the Poynting--Robertson drag
 \citep{10.1098/rsta.1904.0012, 10.1093/mnras/97.6.423, 10.1086/145244,
           10.1016/0019-1035(79)90050-2}.
 For $a_{\textrm{min}}$ we used a fixed value of \mbox{2 $\mu$m}, which is
 smaller than the blow-out grain sizes of our ice-mixture and porous
 silicate grains $a_{\textrm{bo}}^\mathcal{F}$ and $a_{\textrm{bo}}^\mathcal{P}$
 for all values of $\mathcal{F}$ and $\mathcal{P}$ (see Fig.~\ref{fig_bo}). The
 maximum grain size $a_{\textrm{max}}$ has a negligible impact on the appearance
 of a debris disk because the small grains contribute the most to the net
 surface of the disk for the applied steep size distributions. Therefore,
 we set a fixed value of \mbox{$a_{\textrm{max}} = 1$ $\textrm{mm}$}, a value
 often used in observational analyses
 \citep[e.g.,][]{10.1051/0004-6361/201117731, 10.1088/0004-637X/776/2/111,
                  10.1088/0004-637X/792/1/65, 10.1088/0004-637X/798/2/96,
                  10.3847/0004-637X/831/1/97}.
 Eventually, the absolute number of grains as a function of position (see
 Sect.~\ref{debris_disk_model}), grain size, and chemical composition is
 determined by the total dust mass for which we choose an intermediate
 value of \mbox{$M_{\textrm{tot}} = 10^{-8}$ $\textrm{M}_\odot$}
 \citep[e.g.,][and references therein]{10.1086/428348}.

 To determine the blow-out grain size $a_{\textrm{bo}}$ from synthetic
 observations, we need the minimum grain size $a_{\textrm{min}}$ to be a free
 parameter for the reference compact silicate disk sample. We considered the
 interval \mbox{$a_{\textrm{min}} = 0.1$ -- 20 $\mu$m}, sampled by 40
 logarithmically spaced grid values. This interval covers all possible blow-out
 grain sizes in our simulations (see Fig.~\ref{fig_bo}). Furthermore, the
 parameter grid has to remain unaltered by the blow-out of grains by radiation
 pressure. Therefore, we artificially forbid the blow-out of grains by
 radiation pressure for the simulation of the reference disks.


\subsection{Debris disk model and parameter space}\label{debris_disk_model}

 To simulate observable quantities of debris disks, we used an improved version
 of the tool \textbf{D}ebris disks around \textbf{M}ain-sequence \textbf{S}tars
 \citep[\texttt{DMS};][]{10.1051/0004-6361/201833061}. It is based on the
 assumption of an optically thin dust configuration to compute SEDs and
 intensity maps of the thermal dust emission and scattered stellar radiation.

 We considered debris disks in face-on orientation. The dust is distributed in
 a wedge-like structure, restricted by an inner and outer disk radius,
 $R_{\textrm{in}}$ and $R_{\textrm{out}}$, respectively, and a half opening
 angle with a fixed value $\delta = 5 \degr$. The particle number density
 follows a power law $n\left( r \right) \propto r^{-\alpha}$, with $r$ denoting
 the distance from a grain to the central star.

 We varied the inner disk radius $R_{\textrm{in}}$ from $10$ --
 \mbox{$50$ $\textrm{AU}$}, corresponding to the case of cold debris disks.
 Usually, the outer disk radius $R_{\textrm{out}}$ deduced from debris disk
 observations is set by the sensitivity of the observing instrument. We set a
 fixed cutoff at \mbox{$R_{\textrm{out}} = 250$ $\textrm{AU}$}. The exponent
 $\alpha$, describing the radial density profile, usually ranges from values of
 $\alpha = 1$ for a stationary outflow of unbound grains
 \citep{1998A&A...339..477L} to $\alpha = 2.5$ for a transport-dominated disk,
 \citep{10.1086/505736, 10.1051/0004-6361:20064907}.
 \citet{10.1051/0004-6361:20064907} note that a recent large collisional event
 or planets orbiting within the dust distribution could further steepen the
 slope. Therefore, we considered values of $\alpha$ in the range \mbox{1 -- 3}.

 We examined $20$ values for each of the free disk parameters of the
 synthetic observations. They are linearly spaced for the parameters
 $\alpha$ and $\gamma$ (see Sect.~\ref{grain_model}); instead, they are
 logarithmically spaced for the inner disk radius $R_{\textrm{in}}$.
 The parameter space for the reference silicate disks, hence the
 fit parameter space, should be larger than that of the
 synthetic observations to avoid artificial restrictions during
 the fitting process.
 Therefore, we defined the parameter space of the reference models to be
 that of the synthetic observations and further extended it by two values
 into the high and low regime using the same spacing.
 Thus, the fit parameter space contains $24$ values
 for each parameter and is a strict superset of the space of the synthetic
 observations.
 All parameter spaces are summarized in Table~\ref{table_parameter_space}.
\begin{table}
  \caption{Parameter space}
  \label{table_parameter_space}
  \begin{tabular}{l c c c c} 
   \hline \hline
   \addlinespace[2pt]
    Disk & \multicolumn{2}{c}{Synthetic observations} & \multicolumn{2}{c}{Reference models} \\
    parameter & value(s) & n & value(s) & n \\
    \hline 
    \addlinespace[2pt]
    $R_{\textrm{in}} \left[ \textrm{AU} \right]$ &
        10.0 -- 50.0$\,$\tablefootmark{(a)} & 20 & 8.4 -- 59.2$\,$\tablefootmark{(a)}& 24 \\
    $R_{\textrm{out}} \left[ \textrm{AU} \right]$ & 250.0 & 1 & 250.0 & 1 \\
    $\alpha$ & 1.0 -- 3.0$\,$\tablefootmark{(b)}& 20 & 0.8 -- 3.2$\,$\tablefootmark{(b)}& 24 \\
    $\gamma$ & 3.0 -- 4.0$\,$\tablefootmark{(b)}& 20 & 2.9 -- 4.1$\,$\tablefootmark{(b)}& 24 \\
    $a_{\textrm{min}} \left[ \mu \textrm{m} \right]$ &
        $a_{\textrm{bo}}\,$\tablefootmark{(c)} & 1 &
        0.1 -- 20.0$\,$\tablefootmark{(a)}& 40 \\
    $a_{\textrm{max}} \left[ \mu \textrm{m} \right]$ &
        1000.0 & 1 & 1000.0 & 1 \\
    $\delta$ [\degr] & 5.0 & 1 & 5.0 & 1 \\
    $M_{\textrm{tot}} \left[ \textrm{M}_\odot \right]$ &
        $10^{-8}$ & 1 & scaled to obs. & -\\
    $\mathcal{F}$ & 0.1 -- 0.9$\,$\tablefootmark{(b)}& 9 & 0.0 & 1 \\
    \hline
  \end{tabular}
  \tablefoot{
   \tablefoottext{a}{Logarithmically spaced},
   \tablefoottext{b}{Linearly spaced},
   \tablefoottext{c}{Individual for ice-mixture and porous silicate}
            }
  \end{table}


\subsection{Central star model}\label{stellar_model}

 Most of the spatially resolved debris disks are located around
 A-type stars$\,$\footnote{\label{fn_jena_resolved_disks}\href{https://www.astro.uni-jena.de/index.php/theory/catalog-of-resolved-debris-disks.html}{https://www.astro.uni-jena.de/index.php/theory/catalog-of-resolved-debris-disks.html}}.
 As the central star we modelled a star similar to \mbox{$\beta$ Pic}, an A6V
 star with an effective temperature of
 \mbox{$T_\mathrm{eff} = 8052$ $\mathrm{K}$} \citep{10.1086/504637} located at
 a distance of \mbox{$19.44$ $\mathrm{pc}$}\footnote{In the Gaia Early Data
 Release 3 this value has been updated to \mbox{$19.635$ $\mathrm{pc}$}
 \citep{10.1051/0004-6361/201629272, 10.1051/0004-6361/202039657}.} from the
 Solar System \citep{10.1051/0004-6361:20078357}. We applied a stellar mass of
 \mbox{$1.75$ $\textrm{M}_\odot$}
 \citep{1997A&A...320L..29C, 2004IAUS..219...80K} and a stellar radius of
 \mbox{$1.732$ $\textrm{R}_\odot$} \citep{2004IAUS..219...80K}.
 For the stellar spectrum $J_\lambda$ we used a HiRes spectrum from the
 Göttingen Spectral
 Library$\,$\footnote{\href{http://phoenix.astro.physik.uni-goettingen.de}{http://phoenix.astro.physik.uni-goettingen.de}}
 \citep{10.1051/0004-6361/201219058}, calculated for an effective
 temperature and a logarithmic surface gravity of
 \mbox{$T_{\star} = 8000$ $\mathrm{K}$} and
 \mbox{$\log{g} = 4.0\ \left[ \mathrm{cgs} \right]$}, respectively.
 To compute scattered light maps at wavelengths exceeding the maximum
 wavelength of the synthetic spectrum (\mbox{$5.5$ $\mu$m}), the
 stellar spectrum was extrapolated using the Planck function corresponding
 to the effective temperature of the star.

 In Sect.~\ref{subsect_spectral_types} we investigate additional spectral
 types from A0 to K5. The corresponding values of effective temperature and
 logarithmic surface gravity are listed there (see
 Table~\ref{table_stellar_parameters}).


\subsection{Simulated observations}\label{simulated_observations}

 We considered selected observing scenarios in which the choices of wavelengths
 and instrument characteristics are motivated by real debris disk
 observations.
 To mimic the SEDs we selected nine wavelengths ranging from the atmospheric
 $N$ band to \mbox{$1.3$ mm}. These observing wavelengths and the corresponding
 astronomical instruments are typical for existing debris disk observations,
 and have played an important role in advancing the field over the last decades
 (see Table~\ref{table_sed_wavelengths}). To keep the results of this study
 independent of specific instrument characteristics, and thus generally
 applicable, we refrained from applying specific sensitivity limits.
 Furthermore, we did not integrate over the transmission curves of the
 respective instruments and obtained the individual photometric data points for
 distinct wavelengths. Likewise, to keep the study applicable to various debris
 disk systems with different dust masses, we neglected the direct flux of the
 central star to avoid the influence of the specific dust to star flux ratio.
 \begin{table}
  \caption{Considered observing wavelengths}
  \label{table_sed_wavelengths}
  \begin{tabular}{l l} 
   \hline \hline
   \addlinespace[2pt]
   $\lambda \left[ \mu \textrm{m} \right]$  & Instrument \\
   \hline
   \addlinespace[2pt]
   10.5 & e.g., VLT/VISIR$\,$\tablefootmark{(a)} \\
   24 & \textit{Spitzer}$\,$\tablefootmark{(b)}/MIPS$\,$\tablefootmark{(c)} \\
   70 & \textit{Spitzer}/MIPS, \textit{Herschel}$\,$\tablefootmark{(d)}/PACS$\,$\tablefootmark{(e)} \\
   100 & \textit{Herschel}/PACS \\
   160 & \textit{Spitzer}/MIPS, \textit{Herschel}/PACS \\
   214 & SOFIA$\,$\tablefootmark{(f)}/HAWC+$\,$\tablefootmark{(g)} \\
   450 & JCMT/SCUBA2$\,$\tablefootmark{(h)}, ALMA$\,$\tablefootmark{(i)} \\
   850 & JCMT/SCUBA2, ALMA \\
   1300 & ALMA \\
   \hline
  \end{tabular}
  \tablefoot{Typical wavelengths and corresponding astronomical instruments for
             existing debris disk observations.}
  \tablebib{
            \tablefoottext{a}{\citet{2004Msngr.117...12L}}; 
            \tablefoottext{b}{\citet{10.1086/422992}}; 
            \tablefoottext{c}{\citet{10.1086/422717}}; 
            \mbox{\tablefoottext{d}{\citet{10.1051/0004-6361/201014759}};} 
            \tablefoottext{e}{\citet{10.1051/0004-6361/201014535}}; 
            \tablefoottext{f}{\citet{10.1142/S2251171718400111}}; 
            \tablefoottext{g}{\citet{10.1142/S2251171718400081}}; 
            \tablefoottext{h}{\citet{10.1093/mnras/sts612}}; 
            \tablefoottext{i}{\citet{2002Msngr.107....7K}} 
            }

 \end{table}

 To simulate observations of spatially resolved images, we used the
 characteristics of the instruments PACS \citep{10.1051/0004-6361/201014535}
 on \textit{Herschel}, with which the major debris disk surveys DEBRIS
 \citep{10.1051/0004-6361/201014667} and DUNES
 \citep{10.1051/0004-6361/201014594, 10.1051/0004-6361/201321050}
 were performed, and ALMA, the state-of-the-art instrument regarding
 resolving power in the millimeter--submillimeter regime
 \citep[see, e.g.,][]{10.1088/2041-8205/762/2/L21, 10.1126/science.1248726,
                        10.3847/1538-4357/aa71ae, 10.1093/mnras/sty1790}.
 We used two wavelengths for each instrument:
 \mbox{$70$ $\mu$m}, \mbox{$160$ $\mu$m} (\textit{Herschel}/PACS), and
 \mbox{$850$ $\mu$m}, \mbox{$1300$ $\mu$m} (ALMA).
 We convolved the images with an instrument and wavelength-dependent
 circular Gaussian. Subsequently, we extracted radial profiles sampled by
 multiples of the full width at half maximum (FWHM) of the instrument beam,
 starting at the image center (see Appendix~\ref{appdx_sim_images} for a
 detailed description of the simulation of these radial profiles). We assumed
 the disks to always have radial profile data in all four wavelengths.

 In the synthetic observations of ice-containing debris disks we introduced
 artificial uncertainties in the final data products: for each data point we
 set up a normal probability distribution with the precise value of the data
 point as mean and \mbox{$10$ $\%$} of it as standard deviation. Subsequently,
 we produced a new, now final, data point from the probability distribution with
 a random number generator. For the later data analysis (see
 Sect.~\ref{subsect_fitting}) we assumed the final data points of the synthetic
 observations to have an uncertainty value of \mbox{$10$ $\%$}, denoted by
 $\sigma$. Doing so we
 slightly under- or overestimated the uncertainty for data points which became
 lower or higher than their respective true values due to the addition of
 noise. Nonetheless, with this method we obtained data points that scatter
 around their respective true values in a normally distributed fashion.


 \subsection{Fitting approach}\label{subsect_fitting}

 For each of the ice-containing disks we searched for the pure silicate disk
 within the reference sample whose observables mimic most closely those of the
 ice-containing model. We computed the weighted squared deviation $\chi^2$
 and identified the best-fit reference model as the one with the smallest value
 of $\chi^2$. Beforehand, the reference observables have to be scaled with
 the total disk mass $M_\textrm{tot}$ to deviate the least from the synthetic
 ones. This step can be incorporated into the calculation of the weighted
 squared deviation as
 \begin{equation}
  \label{eq_chi_square}
  \chi^2 = \sum_i w_i \left( x_i - f_{M_\textrm{tot}} \hat{x}_i \right)^2 \, ,
 \end{equation}
 with the values of the synthetic observation $x_i$, their weights $w_i$
 (see explanation below), the mass scaling factor $f_{M_\textrm{tot}}$, and
 the values of the reference observation $\hat{x}_i$. To find the mass scaling
 factor $f_{M_\textrm{tot}}$ for which the value of $\chi^2$ is minimal, this
 equation can be differentiated with respect to $f_{M_\textrm{tot}}$ and set to
 zero. The resulting equation can be uniquely solved:
 \begin{equation}
  \label{eq_mass_scaling_factor}
  f_{M_\textrm{tot}}= \frac{\sum_i w_i x_i \hat{x}_i}
                      {\sum_i w_i \hat{x}_i^2} \, .
 \end{equation}
 By combining Equations (\ref{eq_chi_square}) and (\ref{eq_mass_scaling_factor}),
 the value of $\chi^2$ for each model fit is calculated.

 To disentangle the influence of ice on different observables we searched for
 the best-fit model separately using the SED, radial profiles and a combination
 of the two. Regarding weights, we did not consider instrument specific
 uncertainties. For the individual analysis of the SED and radial profiles, the
 weights $w_i$ are defined only by the uncertainties of the data points
 $w_i = \sigma_i^{-2}$. For the combination of the SED and radial profiles we
 used an additional weight causing the two entire data sets to be equally
 weighted, regardless of how many data points they possess individually. The
 SED consists of $9$ data points (see Table~\ref{table_sed_wavelengths}), the
 radial profiles all together consist of $27$ data points
 (see Table~\ref{table_beams_and_rp}), thus
 $w_{i, \, \mathrm{SED}} = 2 \sigma_i^{-2}$ and
 $w_{i, \, \mathrm{RP}} = \frac{2}{3} \sigma_i^{-2}$.

 As a sanity check for our fit routine we used the simulated observables of
 reference silicate disk models themselves as synthetic observations.
 The radial profile fit always results in the correct model parameters, even
 if simulated noise is added. The SED fit does so only without noise; with
 noise the fit results can deviate from the correct parameters because the SED
 is rather insensitive to small changes in disk parameters, and some parameter
 changes are degenerate in their effects on the SED.


\section{Results}\label{sect_results}


\subsection{Reliability of derived debris disk parameters}
\label{subsect_beta_pic_results}

\begin{figure*}
  \centering
  \includegraphics[width=17cm]{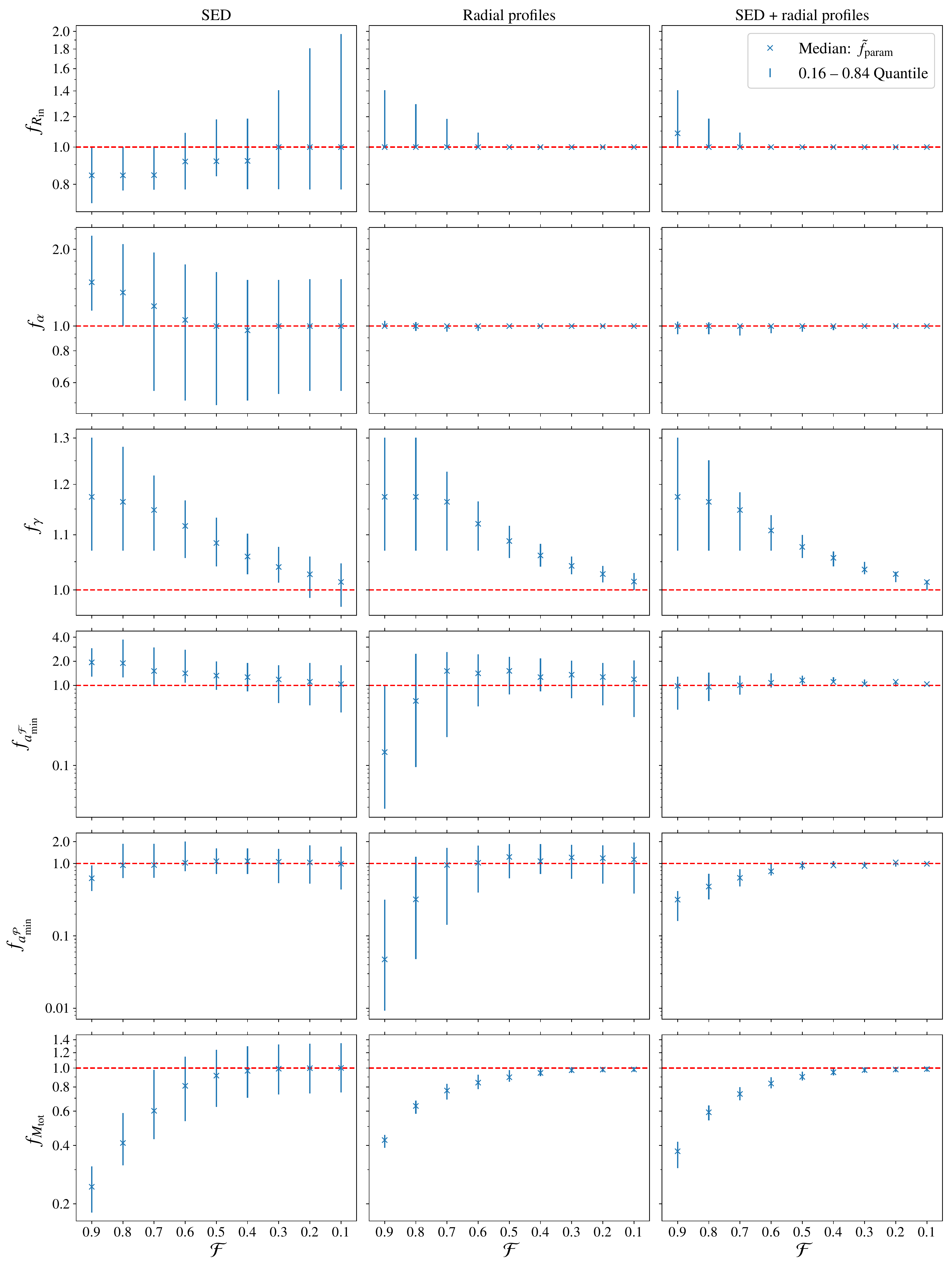}
  \caption{Statistical results of the fitting process of $8000$ model parameter
           sets based on the SED (\textit{left}), the radial profiles
           (\textit{middle}), and the combination of the two (\textit{right}).
           The blue crosses indicate the median factor
           $\tilde{f}_\mathrm{param}$ of all parameter estimations as a
           function of decreasing ice fraction $\mathcal{F}$. For
           $f_\mathrm{param} > 1$ ($< 1$) a given parameter is
           overestimated(underestimated). The dashed red line gives a correct
           estimation with $f_\mathrm{param} = 1$. The vertical lines denote
           the range from the $0.16$ to $0.84$ quantile (i.e., $68\ \%$ of the
           values lie within that interval).}
  \label{fig_stat_results_beta_pic}
 \end{figure*}

 We define $f_\mathrm{param}$ as the ratio of the best-fit value of a given
 parameter to its correct (``true'') value. Thus, we can distinguish between
 the qualitatively different outcomes:
 \begin{equation}
  f_\mathrm{param}\quad
  \begin{cases}
    \textrm{overestimation},& \mathrm{if}\ f_\mathrm{param} > 1\\
    \textrm{correct estimation},& \mathrm{if}\ f_\mathrm{param} = 1\\
    \textrm{underestimation},& \mathrm{if}\ f_\mathrm{param} < 1\; .
  \end{cases}
 \end{equation}
 We obtained $8000$ ($= 20^3$) best-fit parameter sets for each ice fraction
 $\mathcal{F}$, and evaluated them in a statistical manner to identify trends.
 For each value of $\mathcal{F}$ and disk parameter, we derived the median
 $f_\mathrm{param}$, denoted by $\tilde{f}_\mathrm{param}$. Furthermore,
 we derived the $0.16$ and $0.84$ quantiles, defining the interval around
 the median value $\tilde{f}_\mathrm{param}$ containing $68\ \%$ of the values
 of $f_\mathrm{param}$, as a measure of the scatter in the estimated parameter
 values (see Fig.~\ref{fig_stat_results_beta_pic}). In the following we discuss
 the fitting results of the individual parameters.

\subsubsection{Inner disk radius $R_\mathrm{in}$}
 Fitting to the SED, the quantity $R_\textrm{in}$ is preferentially
 underestimated for ice fractions $\mathcal{F} \geq 0.4$. For decreasing ice
 fractions the median value $\tilde{f}_{R_\textrm{in}}$ increases until it
 indicates a match between the fitted and the true value for
 $\mathcal{F} \leq 0.3$. At the same time, the scatter of results and the level
 of overestimation of the inner disk radius $R_\textrm{in}$ increases with
 decreasing value of $\mathcal{F}$.

 Based on the fit to radial profiles, thus incorporating geometrical
 information, we find the median value to be $\tilde{f}_{R_\textrm{in}} = 1$
 and no synthetic observations resulting in an underestimation of the inner
 disk radius within the $68\ \%$ interval for all ice fractions $\mathcal{F}$.
 Nonetheless, for $\mathcal{F} \geq 0.6$ the fraction of models with an
 overestimated inner disk radius increases along with $\mathcal{F}$.
 This can be explained as follows. For disk models with an inner disk radius of
 $\lesssim 25\ \mathrm{AU}$, the temperature of grains with radii of
 \mbox{$\sim$ 5 $\mu$m} up to several hundred \mbox{$\mu$m} exceeds the ice
 sublimation temperature inside \mbox{$\sim$ 20 -- 25 AU} (exact values
 dependent on the ice fraction $\mathcal{F}$). For all four wavelengths at
 which we investigate radial profiles (see Table~\ref{table_beams_and_rp}), it
 holds that the remaining porous grains emit less thermal radiation in the
 respective wavelength than ice-mixture grains would, leading to a pronounced
 decrease in the flux level. The flux deficit increases with wavelength and ice
 fraction $\mathcal{F}$. The fitting process, based on the reference model with
 a smooth flux radial profile, favors configurations with a larger inner disk
 radius to adapt to the flux deficit in close-in regions.

 When comparing the results of combining the SED and radial profiles in
 a single fit to those of the pure radial profile fit, we see that the scatter
 in results is reduced for $\mathcal{F} \in \left\{ 0.8, 0.7, 0.6 \right\}$,
 but not for $\mathcal{F} = 0.9$.

\subsubsection{Radial density distribution exponent $\alpha$}
 Fitting to the SED, the radial density decrease is mainly overestimated for
 ice fractions of $\mathcal{F} \geq 0.6$. For smaller ice fractions the
 estimations are better; for $\mathcal{F} \leq 0.5$ the median value
 $\tilde{f}_\alpha = 1$ except for $\mathcal{F} = 0.4$, but there the deviation
 from one is small. The scatter in results is generally large, even for the
 lowest values of $\mathcal{F}$. As expected for the determination of a
 geometrical parameter, the fit to radial profiles performs better; we find
 $\tilde{f}_\alpha = 1$ for all ice fractions $\mathcal{F}$, with only a small
 scatter in results for $\mathcal{F} \geq 0.6$. When combining SED and radial
 profile information, the scatter in results increases slightly for
 $\mathcal{F} \geq 0.4$.

\subsubsection{Grain size distribution exponent $\gamma$}
 We see the same trend in the results of fitting to the SED and fitting to the
 radial profiles.
 Except for the SED case with low ice fractions
 $\mathcal{F} \in \left\{ 0.2, 0.1 \right\}$, $\gamma$ gets
 strictly overestimated. The mismatch is the largest with a median value
 of $\tilde{f}_\gamma \sim 1.2$ for the highest ice fraction
 $\mathcal{F} = 0.9$ and decreases along with it.
 Therefore, for a higher value of $\mathcal{F}$ a higher fraction of small
 grains is needed to reproduce the synthetic observations.
 The scatter in results for ice fractions $\mathcal{F} \geq 0.7$ is almost
 the same for the fit to the SED and to radial profiles; for lower
 ice fractions $\mathcal{F} \leq 0.6$ fitting to the radial profiles results
 in slightly smaller scattering.

 The above trend is similar to the one \citet{10.1093/mnras/stw2675} find when
 investigating the effect of porous silicate grains. At first glance it
 appears, that this finding can be explained by the presence of porous
 silicate in the inner disk regions where ice is sublimated. However, this
 explanation does not imply that the impact of ice is negligible.
 First, when artificially suppressing ice sublimation during the simulation,
 the fitting results for the exponent of the grain size distribution $\gamma$
 remains almost unaffected. The same is true in the case of stars with later
 spectral types (see Sect.~\ref{subsect_spectral_types}), which are not
 luminous enough to sublimate significant amounts of ice. Thus, we conclude
 that the above trend has its origin in the substitution of some amount of
 silicate by an optical medium that is less optically active (in the respective
 wavelength range) and dense, be it a vacuum or ice. Therefore, the fit using
 pure silicate as the grain material assumes the grains to be too massive. This
 explanation coincides with our finding that ice-mixture grains lead to a
 smaller overestimation of the grain size distribution exponent $\gamma$ than
 porous silicate grains do
 \citep[see lower right panel of Fig.~3 in][]{10.1093/mnras/stw2675}
 because ice is optically active and massive, while a vacuum is neither.

\subsubsection{Minimum grain size $a_\textrm{min}$}
\label{subsubsect_a_min_results}
 From the fit we obtained a single estimation of the minimum grain size
 $a_\textrm{min}$.
 First we compared it to the blow-out grain size of the ice-mixture
 $a_{\textrm{bo}}^\mathcal{F}$ and derived the factor
 $f_{a_{\textrm{min}}^\mathcal{F}}$. As shown in Fig.~\ref{fig_bo}, ice-mixture
 grains of the size $a_{\textrm{bo}}^\mathcal{F}$ are the smallest grains
 in the disks as $a_{\textrm{bo}}^\mathcal{F} <
 a_{\textrm{bo}}^\mathcal{P}$ for all ice fractions $\mathcal{F}$.
 In the median the results from fitting to the SED overestimate the parameter
 $a_\textrm{min}$ for all values of $\mathcal{F}$, at most by a factor
 $\tilde{f}_{a_{\textrm{min}}^\mathcal{F}} \sim 2$ for $\mathcal{F} = 0.9$.
 While $\tilde{f}_{a_{\textrm{min}}^\mathcal{F}}$ decreases along with
 the value of $\mathcal{F}$, the scatter of results does not. The results of
 the fit to radial profiles for ice fractions $\mathcal{F} \leq 0.7$ are
 similar to those of the fit to the SED. However, for large ice fractions
 $\mathcal{F} \geq 0.8$ they show a completely different behavior; they
 underestimate the minimum grain size $a_{\textrm{min}}^\mathcal{F}$ with a
 very large scatter. The possibility that the fit favors a smaller minimum
 grain size $a_{\mathrm{min}}$ to compensate the overestimation of the inner
 disk radius $R_\mathrm{in}$ can be ruled out. In a simulation with suppressed
 sublimation the fitting results for $R_\mathrm{in}$ scatter less for a given
 ice fraction $\mathcal{F}$, while those of $a_{\textrm{min}}^\mathcal{F}$ are
 barely affected. The behavior for $\mathcal{F} \geq 0.8$ can be explained by
 the fact that high ice fractions $\mathcal{F}$ cause the radial profiles of
 the ice-containing disks to become flatter compared to those for lower ice
 fractions. The fit routine, working only with pure silicate disks, cannot vary
 the dust material to adapt to this. However, within the chosen model setup
 (i.e., defined by the properties of the central star, disk geometry, observing
 wavelengths, dust material), decreasing the minimum grain size
 $a_\textrm{min}$ in the silicate disks has the same effect as decreasing the
 slope of the radial profiles. Therefore, to fit the radial profiles of the
 ice-containing disks that possess the highest ice fractions
 $\left( \mathcal{F} \geq 0.8 \right)$ the fit is biased toward smaller values
 of $a_\textrm{min}$, that is, an underestimation of that parameter.

 Contrary to the determination of all other disk parameters, here the
 combination of the two data sets, SED and radial profiles, has a major
 effect. Compared to the fit results of the single data sets the scatter
 of parameter estimations decreases for all ice fractions $\mathcal{F}$ and
 the median value $\tilde{f}_{a_{\textrm{min}}^\mathcal{F}} \approx 1$
 even for the highest values of $\mathcal{F}$.

 Then we compared the estimated minimum grain size $a_\textrm{min}$ to the
 blow-out grain size of porous silicate $a_{\textrm{bo}}^\mathcal{P}$
 and obtained $\tilde{f}_{a_{\textrm{min}}^\mathcal{P}}$.
 We find that the fit to the SED for ice fractions of $\mathcal{F} \leq 0.8$
 and the fit to radial profiles for $\mathcal{F} \leq 0.7$ are biased toward
 the size of the smallest porous grains, instead of favoring the size of the
 smallest ice-mixture grains.
 For the fit to the combined data this is only the case for
 ice fractions of $\mathcal{F} \leq 0.5$, where the difference between
 the two blow-out grain sizes $a_{\textrm{bo}}^\mathcal{F}$ and
 $a_{\textrm{bo}}^\mathcal{P}$ is already small (see Fig.~\ref{fig_bo}).

\subsubsection{Total disk mass $M_\mathrm{tot}$}
 A clear trend with ice fraction $\mathcal{F}$ is apparent for the results from
 fitting to the SED and to the radial profiles. While the median value
 $\tilde{f}_{M_\mathrm{tot}} \approx 1$
 for ice fractions $\mathcal{F} \leq 0.4$, a further increase in $\mathcal{F}$
 results in an underestimation of the disk mass $M_\mathrm{tot}$. Thus, the
 reference disks require a lower mass than the ice-containing ones to reproduce
 a similar flux. Here, three effects are important. The first causes an
 underestimation of the
total disk mass, while the second and third cause an overestimation,
 effectively canceling each other out. The higher the ice fraction
 $\mathcal{F}$, respectively porosity $\mathcal{P}$, the stronger all three
 effects are.

 First, as the compact silicate grains have a higher mass than equally sized
 ice-mixture and porous silicate grains, a disk consisting of compact silicate
 grains requires more mass to possess the same number of grains, thus the same
 total cross section, as a disk consisting of the more lightweight ice-mixture
 or porous silicate grains. The discrepancy increases with increasing
 ice fraction $\mathcal{F}$ and porosity $\mathcal{P}$.

 Second, the higher the value of $\mathcal{F}$, the higher the fitting
 results for the quantity $\gamma$, hence the steeper the grain size
 distribution (see Fig.~\ref{fig_stat_results_beta_pic} and earlier in this
 section). While the small grains contribute only a minor fraction to the disk
 mass, they contribute the most to the total cross section, thus to the emitted
 and scattered radiation.
 A disk with a steeper size distribution possesses a higher fraction of small
 grains than a disk with a shallower size distribution. Therefore, to produce
 the same flux level a disk with a steeper size distribution requires less mass
 than a disk with a shallower distribution.

 Third, as compact silicate grains are hotter
 and emit more thermal radiation than ice-mixture and porous grains at the same
 radial distance from the central star, a disk consisting of compact silicate
 grains requires fewer grains, thus less mass, to produce the same flux level
 as a disk consisting of ice-mixture or porous silicate grains.

 For ice fractions of $\mathcal{F} \leq 0.4$ the effects compensate for each
 other, leading to a correct estimation of the total disk mass. However,
 toward larger values of $\mathcal{F}$ the latter two effects dominate and the
 fitting results underestimate the disk mass. The degree of underestimation
 from fitting to the SED is significantly stronger than that from fitting to
 radial profiles, for instance, with $\tilde{f}_{M_{\textrm{tot}}} \sim 0.25$
 for the fit to SED and $\tilde{f}_{M_{\textrm{tot}}} \sim 0.4$ for the fit to
 radial profiles for an ice fraction of $\mathcal{F} = 0.9$.
 Furthermore, the scatter of fitting results based on the SED is generally
 larger than that based on the radial profiles, even for low ice fractions.
 Combining the two data sets increases the scatter of the results from fitting
 solely to radial profiles.

\subsubsection{General evaluation}
 The results from fitting to the SED are degenerate for ice fractions of
 $\mathcal{F} \gtrsim 0.5$. The inner disk radius $R_\mathrm{in}$ is
 underestimated, while at the same time the slopes of the radial density
 distribution $\alpha$ and the grain size distribution $\gamma$ are
 overestimated. A smaller value of $R_\mathrm{in}$ allows grains closer to the
 central star, a larger value of $\alpha$ increases the grain number density in
 those close-in regions, and a larger value of $\gamma$ increases the amount of
 small grains compared to larger grains. Thus, the SED fitting favors disk
 models with smaller and hotter grains compared to the ice-containing disks to
 reproduce their SEDs. However, this effect is balanced by an
 overestimation of the minimum grain size $a_\mathrm{min}$, which sets the
 lower limit of the grain size distribution. Therefore, the SED fit results
 are degenerate between more grains in regions where they are hotter together
 with an increase in the abundance of smaller grains, and limiting the grain
 size distribution at the lower end.

 Adding SED information to the radial profiles is only of major use for the
 determination of the minimum grain size $a_\mathrm{min}$.
 For all other debris disk parameters investigated the benefit is small
 or the scatter of results even increases. This behavior is dependent on the
 quality of the data (number of data points, noise level) and the chosen
 weighting of SED against radial profiles. For a worse quality of SED data,
 adding it to the radial profiles can be detrimental. Thus, when high-quality
 radial profiles are available, it can be advantageous to use SED data
 only when determining the respective disk parameters.


\subsection{Different distances to the debris disk system}
\label{subsect_dist_variation}

\begin{figure}
  \centering
  \resizebox{\hsize}{!}{\includegraphics{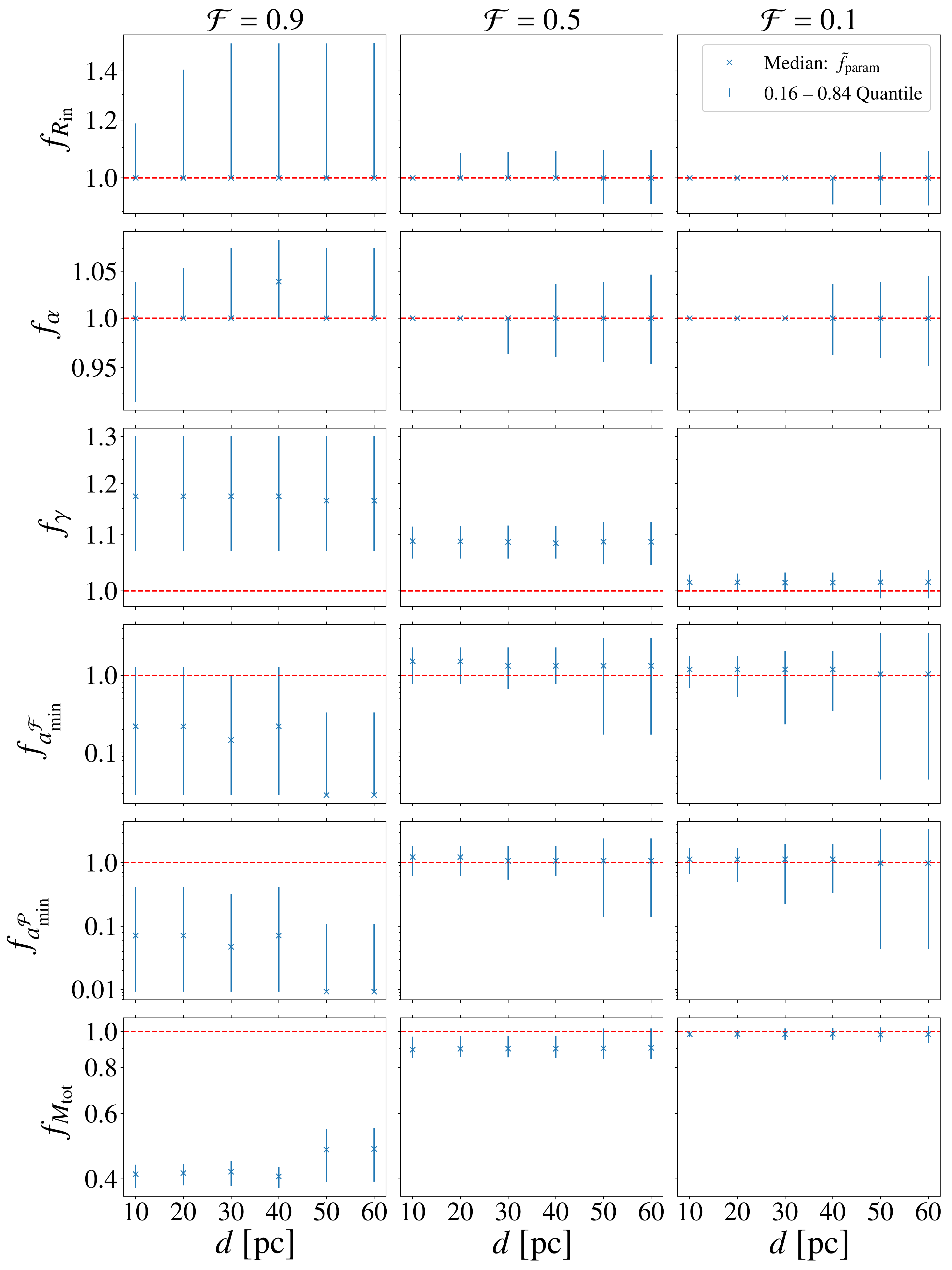}}
  \caption{Statistical results of the process of fitting to radial profiles
           for different distances to the debris disk system. Shown are the
           results for the three ice fractions $\mathcal{F} = 0.9$
           (\textit{left column}), $0.5$ (\textit{middle column}), and $0.1$
           (\textit{right column}). The results are presented as in
           Fig.~\ref{fig_stat_results_beta_pic}.}
  \label{fig_dist_study_beta_pic}
 \end{figure}

 To date debris disks have been spatially resolved at different distances
 up to \mbox{$\sim 150\ $pc;} roughly two-thirds are within
 \mbox{$60\ $pc}\footref{fn_jena_resolved_disks}. We now investigate the
 influence of the distance to the debris disk and thus the achievable spatial
 resolution on the parameter estimations. With the same setting as before, we
 set the distance to six different values from \mbox{$10\ $pc} to
 \mbox{$60\ $pc}. Along with the increasing distance to the debris disk system
 the spatial resolution of the images is decreasing and we extracted fewer data
 points for the radial profiles (see Sect.~\ref{simulated_observations} and
 \ref{appdx_sim_images}). Due to the limited sensitivity of a given observing
 instrument, the signal-to-noise ratio depends on the brightness of a specific
 object. However, specific instrument characteristics are not considered in our
 analysis. Therefore, the analysis of our synthetic SEDs is not affected by
 the distance variation and we restrict our discussion here to the results of
 fitting to radial profiles (see Fig.~\ref{fig_dist_study_beta_pic}).

 In general, the scatter of results increases along with increasing distance,
 and thus decreasing spatial resolution. This occurs because with fewer data
 points the influence of the noise on the fit results increases. An exception
 is the result for $\alpha$ with $\mathcal{F} = 0.9$ and \mbox{$d = 10$ pc}.
 Likewise, noteworthy are the results for the minimum grain size
 $a_\mathrm{min}$ with distances of \mbox{$d \geq 50$ pc}. For the ice
 fractions $\mathcal{F} \in \left\{ 0.5, 0.1 \right\}$ the scatter of results
 is larger, while for $\mathcal{F} = 0.9$ the median estimations
 $\tilde{f}_{a_{\textrm{min}}^\mathcal{F}}$,
 $\tilde{f}_{a_{\textrm{min}}^\mathcal{P}}$ are smaller by one magnitude than
 those for distances \mbox{$d \leq 40$ pc}. Furthermore, the median values
 correspond to the smallest grain size in the fit parameter space. We
 conclude that the determination of the minimum grain size $a_\mathrm{min}$
 is more sensitive to the spatial resolution of the radial profiles than the
 other disk parameters discussed.
 This behavior also translates into an increase in the total disk mass
 $M_\mathrm{tot}$ determined for an ice fraction of $\mathcal{F} = 0.9$
 and a distance of \mbox{$d \geq 50$ pc}. However, the estimation of the
 grain size distribution exponent $\gamma$ is hardly affected. This coincides
 well with the fact that the fit to the SEDs, which can be seen as completely
 unresolved images, and the fit to radial profiles show similar results for
 this parameter (see Fig.~\ref{fig_stat_results_beta_pic}).


 \subsection{Different stellar types}\label{subsect_spectral_types}

 In addition to A-type stars, debris disks have been observed and resolved
 around various different stellar types. The central star has a major impact on
 its debris disk. It affects the dust temperature distribution, and thus the
 thermal emission of the grains, as well as the radial location beyond which
 ice can survive. Furthermore, the strength of the gravitational field paired
 with the radiation pressure changes the blow-out grain size $a_{\mathrm{bo}}$,
 allowing grains of different sizes within the system.

 We performed the same investigation as for the \mbox{$\beta$ Pic} sibling for
 a set of main-sequence stellar types from A0 to K5. The stellar properties
 were taken from \citet{1998gaas.book.....B}. The simulated stars and their
 parameters are listed in Table~\ref{table_stellar_parameters}. The
 corresponding values of stellar effective temperature $T_{\star}$ and
 logarithmic surface gravity $\log g$ are chosen as the nearest grid values in
 the Göttingen spectral library.
 \begin{table}
  \caption{Stellar parameters of main-sequence stars}
  \label{table_stellar_parameters}
  \begin{tabular}{l l l l l}
   \hline \hline
   \addlinespace[2pt]
   Spectral type & $T_{\star} \left[ \mathrm{K} \right]$ &
   $R_{\star} \left[ \mathrm{R_\odot} \right]$ &
   $M_{\star} \left[ \mathrm{M_\odot} \right]$ &
   $\log g \left[ \mathrm{cgs} \right]$ \\
    \hline
    \addlinespace[2pt]
    A0 & 9600 & 2.4 & 2.9 & 4.0 \\
    A5 & 8200 & 1.7 & 2.0 & 4.5 \\
    F0 & 7200 & 1.5 & 1.6 & 4.5 \\
    F5 & 6400 & 1.3 & 1.3 & 4.5 \\
    G0 & 6000 & 1.1 & 1.05 & 4.5 \\
    G5 & 5800 & 0.92 & 0.92 & 4.5 \\
    K0 & 5300 & 0.85 & 0.79 & 4.5 \\
    K5 & 4400 & 0.72 & 0.67 & 4.5 \\
   \hline
  \end{tabular}
  \tablefoot{Values of $T_{\star}$ and $\log g$ given by the nearest grid
             values in the Göttingen spectral library.}

  \tablebibsingular{\citet{1998gaas.book.....B}, Tables 3.7 and 3.13}
 \end{table}

 Due to the small radiation pressure, for central stars with spectral types
 of F0 and later either the blow-out grain size
 $a_\textrm{bo} < 1\ \mu \textrm{m}$ or there is no blow-out of grains at all.
 In an exploratory study \citet{10.1051/0004-6361/201935341} investigated
 the presence and impact of those submicrometer grains in a cold
 debris disk composed of compact silicate around an A6V and a G2V star.
 They find for the latter that grains of the size $a \leq a_{\mathrm{bo}}$
 can account for more than \mbox{90\ \%} of the dust cross section.
 Apart from that there is little known about the fate of submicrometer grains
 around late-type stars \citep{10.2458/azu_uapress_9780816531240-ch023}.
 However, except for the two shortest wavelengths in our synthetic SED,
 the relevance of such small grains in our study should be small. Therefore,
 for the ice-containing disks around F0 and later spectral types we set a
 lower border of the grain size distribution of
 $a_\mathrm{min} = 1\ \mu \textrm{m}$. Furthermore, for these spectral types we
 also increased the maximum value of the fit parameter space of the minimum
 grain size $a_\mathrm{min}$ to \mbox{$30\ \mu$m}.

 Comparing the A5 and $\beta$ Pic (A6) star, all estimations of disk parameters
 are very similar. Regarding the fit to radial profiles, for the A0 star a
 larger fraction of inner disk radii $R_\mathrm{in}$ are overestimated, and the
 overestimation is larger than for the A5 and $\beta$ Pic star. Toward later
 spectral types the effect weakens for F5, and later it disappears, and the
 parameter $R_\mathrm{in}$ is estimated reliably by fitting to radial profiles
 for all ice fractions. This is as expected because the later the spectral
 type, the more the ice sublimation radii move inward until all grains retain
 their ice. Except for the A5 star, adding SED information has a negligible
 impact on the results; in some cases it can even increase the scatter of the
 distribution of the derived parameter values.

 The exponent of the radial density distribution $\alpha$ can be constrained
 reliably by radial profiles regardless of the ice fraction $\mathcal{F}$ and
 spectral type. The only exception arises for the A0 spectral type and
 ice fractions $\mathcal{F} \geq 0.5$. In this case the quantity $\alpha$ can
 be slightly underestimated, the strongest with a median factor of
 $\tilde{f}_\alpha \sim 0.8$ for $\mathcal{F} \geq 0.9$. This is again due to
 the sublimation of ice. The bright central star produces extended regions of
 porous silicate grains ranging outward, while in the outer disk regions the
 ice can still survive. As the porous silicate grains have a lower emissivity,
 the fit favors an overall flatter shape of the radial profile. Incorporating
 the SED data does not help to ease this problem.

 In the results of all three fitting approaches the trends for the parameters
 $\gamma$ and $M_\mathrm{tot}$ as seen for the $\beta$ Pic case are almost
 constant with spectral type with only minor variances. Thus, the determination
 of these quantities is mostly independent from the amount of sublimated ice in
 the disk, hence the amount of porous silicate grains, and independent from the
 illuminating star.

\begin{figure}
 \centering
 \resizebox{\hsize}{!}{\includegraphics{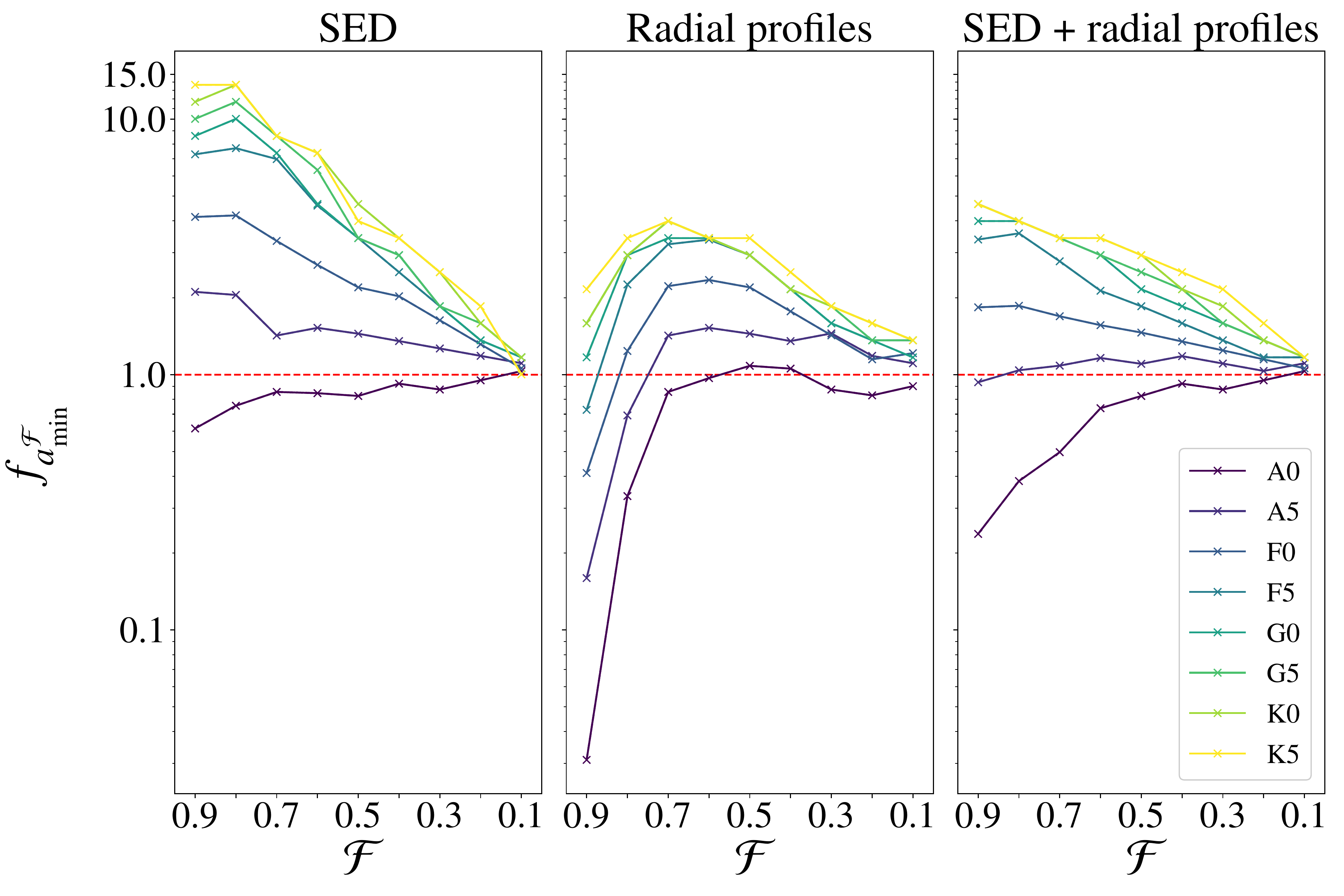}}
 \caption{Statistical results of the estimation of the minimum ice-mixture
          grain size $a_{\textrm{min}}^\mathcal{F}$ for different main-sequence
          spectral types. The quantity $a_{\textrm{min}}^\mathcal{F}$ is the
          smallest grain size in the disk. For the spectral types A0 and A5 it
          is set by the blow-out grain size $a_\mathrm{bo}$; for all later
          types it is fixed to \mbox{$a_{\mathrm{min}} = 1\ \mu \mathrm{m}$}.
          In contrast to the presentation in
          Fig.~\ref{fig_stat_results_beta_pic}, the $0.16$ -- $0.84$ quantile
          intervals are omitted for the sake of clarity.}
 \label{fig_a_min_spectral_type}
 \end{figure}

 Contrary to the other parameters, the determination of the minimum grain size
 $a_\mathrm{min}$ is strongly influenced by the properties of the central star.
 In Fig.~\ref{fig_a_min_spectral_type} the results of the estimations of
 $a_{\textrm{min}}^\mathcal{F}$ for the different spectral types are displayed.
 As stated previously, $a_{\textrm{min}}^\mathcal{F}$ is the smallest grain
 size in the disks; for the A0 and A5 case it is set by $a_\mathrm{bo}$, and for
 later spectral types it is fixed to \mbox{$1\ \mu \mathrm{m}$}. Apart from a
 few exceptions, the later the spectral type, the larger the median value
 $\tilde{f}_{a_{\textrm{min}}^\mathcal{F}}$. Around F5 and later-type
 stars the results converge to a similar distribution.

 Regarding fitting to the SED (see Fig.~\ref{fig_a_min_spectral_type}
 \textit{left}), the minimum grain size $a_\mathrm{min}$ is
 underestimated for an A0 star and ice fractions $\mathcal{F} \geq 0.2$,
 while around an A5 and later-type stars it is overestimated.
 For stars of spectral type F5 and later and an intermediate ice fraction of
 $\mathcal{F} = 0.5$, the median overestimation of $a_\mathrm{min}$ is as
 high as $\tilde{f}_{a_{\textrm{min}}^\mathcal{F}} \sim 3$ -- $5$.
 For the highest ice fractions $\mathcal{F} \geq 0.8$ the median values
 increase up to $\tilde{f}_{a_{\textrm{min}}^\mathcal{F}} \sim 7$ -- $14$.
 These large overestimations can partly be explained by the underestimation of
 the inner disk radius $R_\mathrm{in}$, which results in an increase in the
 abundance of hot grains, an effect which is relaxed in the case of larger
 grains. Although fitting to the radial profiles allows the
 inner disk radius $R_\mathrm{in}$ to be determined reliably, the same general
 trend with spectral type is present for the determination of the minimum grain
 size $a_\mathrm{min}$ as for the results of fitting to the SED (see
 Fig.~\ref{fig_a_min_spectral_type} \textit{middle}). Nonetheless, the results
 of radial profile fitting show a different trend with ice fraction
 $\mathcal{F}$. With increasing $\mathcal{F}$ the median estimation
 $\tilde{f}_{a_{\textrm{min}}^\mathcal{F}}$ increases to a maximum for
 ice fractions $\mathcal{F} \sim 0.4$ -- $0.7$ and then drops sharply.
 The highest overestimations occur for spectral types of F5 and later;
 they are in the median
 $\tilde{f}_{a_{\textrm{min}}^\mathcal{F}} \sim 3$ -- $4$ (see
 Sect.~\ref{subsubsect_a_min_results} for parameters influencing the estimation
 of the minimum grain size $a_{\textrm{min}}$).

 The results of the combined fit (see Fig.~\ref{fig_a_min_spectral_type}
 \textit{right}) are between the results of the SED and radial profiles. We
 find the same clear trend with spectral type toward an overestimation of the
 minimum grain size $a_\mathrm{min}$.


\section{Conclusions}\label{sect_discussion}

 We investigated the effect of water ice on the quantitative analysis of debris
 disk observations. Based on the simulation of selected typical observable
 quantities of a large set of ice-containing debris disks around a star similar
 to \mbox{$\beta$ Pic,} we derived the values of various parameters describing
 these disks. While the simulated observations were performed for debris disks
 with an ice fraction of 0.1 to 0.9, compact silicate dust grains were assumed
 in the analysis of those observations.

 To model the icy dust material we employed EMT mixtures of ice and silicate.
 Ice sublimation was considered by replacing an ice-mixture grain with a porous
 silicate grain if its temperature exceeded the ice sublimation temperature.
 We analyzed SEDs, radial profiles, and a combination of the two.
 Furthermore, we investigated the influence of the distance to the debris disk
 system, and thus the impact of the achievable spatial resolution for fixed
 instrument characteristics, that is, angular resolution. We found that when
 fitting radial profiles, for the debris disk parameters inner radius
 $R_\mathrm{in}$, exponent of the radial density distribution $\alpha$, and
 minimum grain size $a_{\mathrm{min}}$ the reliability of the derived parameter
 values degrades with decreasing spatial resolution, while the exponent of the
 grain size distribution $\gamma$ and the disk mass $M_{\mathrm{tot}}$ are
 barely affected. Lastly, we examined the influence of the stellar spectral
 type (A0 to K5) on the analysis.

 We found that by using radial profiles the inner disk radius $R_\mathrm{in}$
 and the exponent of the radial density distribution $\alpha$ can be
 constrained well, regardless of the ice fraction $\mathcal{F}$ and spectral
 type. However, there is one exception. For the grains covering a large size
 range (\mbox{$a \sim$ 5 $\mu$m} up to several hundred \mbox{$\mu$m}), the
 sublimation of ice occurs in a rather small radial range, almost independent
 of the ice fraction $\mathcal{F}$, but mainly determined by the spectral type
 of the central star. If this radial range is located inside the dust
 distribution, we found an overestimation of the the inner disk radius
 $R_\mathrm{in}$. For very luminous stars this also causes the slope of the
 radial density distribution $\alpha$ being determined too flat. We found this
 only for the A0-type central star. Therefore, if dust grains are expected to
 be icy in a given debris disk, for the analysis of observations a dust model
 is needed that can distinguish between grains in inner disk regions without
 ice and grains in outer disk regions with ice. Otherwise, the size of the
 inner disk cavity is overestimated. This can alter the implications for
 possible stellar companions orbiting within that cavity
 \citep[e.g.,][and references therein]{10.1088/1674-4527/10/5/001}.
 By artificially suppressing ice sublimation we found that the bias in the
 estimation of the parameters $R_\mathrm{in}$ and $\alpha$ has only a minor
 effect on the derived values of the other parameters.

 The exponent of the grain size distribution $\gamma$ is overestimated
 with increasing ice fraction $\mathcal{F}$ by both SED and radial
 profile fitting, independent of the spectral type. This is likewise found
 by \citet{10.1093/mnras/stw2675}. We conclude that the behavior is solely
 caused by assuming the grains in the reference disks to be too massive. Our
 results are in agreement with other methods to determine the
 grain size distribution exponent. For example,
 \citet{10.1051/0004-6361/202037858} systematically investigated the method of
 deriving $\gamma$ from the spectral index at millimeter photometric
 measurements between \mbox{$\lambda \sim 1\ $mm} and \mbox{$7$ -- $9$ mm},
 that is, at longer wavelengths than we considered in our study. He finds
 larger values of $\gamma$ when using silicate than when using pure
 crystalline water ice. The grain size distribution exponent $\gamma$ is
 linked to the critical energy for fragmentation and dispersal
 $Q_\mathrm{D}^\star$ \citep{10.2458/azu_uapress_9780816531240-ch023}. For the
 highest value of $\mathcal{F}$ we found median overestimations by a factor of
 almost $1.2$. Such an overestimation of $\gamma$ could for instance lead to a
 critical energy $Q_\mathrm{D}^\star$ in the gravity regime ($\gamma \sim 3.0$)
 being determined as being in the strength regime ($\gamma \sim 3.6$).

 Within the considered grain size interval the resulting disk mass also does
 not depend on the spectral type. We found an underestimation with
 increasing ice fraction $\mathcal{F}$ which is stronger when fitting to the
 SED as to radial profiles. While for low values of $\mathcal{F}$ the median
 estimation is reliable, the level of underestimation increases along with
 $\mathcal{F}$ up to a factor of $\sim 0.25$ for the SED fit and $\sim 0.4$ for
 both, the radial profile and the combined fit.

 The derived minimum grain size $a_\mathrm{min}$ strongly depends on the
 spectral type, particularly toward late spectral types: the later the spectral
 type, the higher the overestimation. The trend is found both for SED and
 radial profile fitting, but is particularly strong for the former. Regarding
 SED fitting, analyzing the SEDs of $34$ \textit{Herschel} resolved debris
 disks using compact silicate as dust material,
 \citet{10.1088/0004-637X/792/1/65} find the same trend. For some solar-like
 stars the blow-out grain size is overestimated by a factor of up to
 $a_\mathrm{min}$/$a_\mathrm{bo} \sim 10$. In addition to other possible
 explanations, they propose that this trend could be caused by water ice making
 up parts of the dust material. With ice fractions of $\mathcal{F} \gtrsim 0.7$
 we find such large overestimations in our study, and can confirm their
 assumption. Likewise, assuming an ice-mixture similar to ours with
 $\mathcal{F} = 0.5$, \citet{10.3847/0004-637X/831/1/97} find a similar trend
 by analyzing a sample of A-type and solar-like stars. For the solar-like stars
 they find \mbox{$a_\mathrm{min}/a_\mathrm{bo} \sim 5$ -- $6$}, while for the
 A-type stars \mbox{$a_\mathrm{min}/a_\mathrm{bo} \sim 1$}. From our results a
 larger ice fraction $\mathcal{F} > 0.5$ would decrease the ratio they find for
 the solar-like stars. However, water ice is likely not the only solution
 to the mismatch between the values of the minimum grain size $a_\mathrm{min}$
 determined from observations on the one hand and the blow-out grain size
 $a_\mathrm{bo}$ determined on the basis of simple dust/star models on the
 other hand. Other factors discussed so far are dust grain porosity
 \citep{10.1051/0004-6361/201220486, 10.1093/mnras/stv2142,
        10.1093/mnras/stw2675},
 inequalities in the dust production and destruction rates
 \citep{10.1051/0004-6361:20079133}, and the surface energy constraint
 \citep{10.1051/0004-6361/201423862, 10.1093/mnras/stv2142,
        10.1051/0004-6361/201527626}.

 Our study is based on synthetic observations at wavelengths from
 $10$ -- \mbox{$1300\ \mu$m} where the thermal emission of the dust grains
 dominates the radiation. At shorter wavelengths toward the optical to
 near-infrared wavelength range the relative importance of the stellar light
 scattered by the dust grains increases. Due to the anisotropy of the
 scattering function, the disk inclination becomes another parameter that would
 have to be considered in the analysis. In addition, the direct stellar
 radiation complicates the analysis unless the debris disk can be spatially
 resolved and clearly be distinguished from the central star. Nonetheless,
 there are instruments already providing us with spatially resolved images in
 the respective wavelengths such as the HST, GPI, SCExAO, and SPHERE, and new
 instruments with increased sensitivity and angular resolving capabilities are
 expected to become available, such as the \textit{Extremely Large Telescope}
 (ELT) and the JWST.


\section*{ORCID iDs}
T. A. Stuber \orcidlink{0000-0003-2185-0525}
\href{https://orcid.org/0000-0003-2185-0525}
     {https://orcid.org/0000-0003-2185-0525}\\
S. Wolf \orcidlink{0000-0001-7841-3452}
\href{https://orcid.org/0000-0001-7841-3452}
     {https://orcid.org/0000-0001-7841-3452}


\begin{acknowledgements}
 The authors thank the anonymous referee for his comments and suggestions that
 helped to improve the presentation of the results.
 This research has made use of NASA's Astrophysics Data System Bibliographic
 Services, \texttt{adstex}$\,$\footnote{\href{https://github.com/yymao/adstex}{https://github.com/yymao/adstex}},
 a modified A\&A bibliography style
 file$\,$\footnote{\href{https://github.com/yangcht/AA-bibstyle-with-hyperlink}{https://github.com/yangcht/AA-bibstyle-with-hyperlink}},
 Ipython \citep{10.1109/MCSE.2007.53}, Jupyter notebooks \citep{jupyter},
 \texttt{Astropy}\footnote{\href{https://www.astropy.org}{https://www.astropy.org}},
 a community-developed core Python package for
 Astronomy \citep{10.1051/0004-6361/201322068, 10.3847/1538-3881/aabc4f},
 Matplotlib \citep{10.1109/MCSE.2007.55} and Numpy
 \citep{10.1038/s41586-020-2649-2}.
 This work was supported by the Research Unit FOR 2285 “Debris Disks in
 Planetary Systems” of the Deutsche Forschungsgemeinschaft (DFG). The authors
 acknowledge the DFG for financial support under contract WO 857/15-2.
\end{acknowledgements}


\bibliographystyle{aa_url} 
\bibliography{bibliography.bib}

\begin{thebibliography}{93}
\expandafter\ifx\csname natexlab\endcsname\relax\def\natexlab#1{#1}\fi

\bibitem[{{Astropy Collaboration} {et~al.}(2018){Astropy Collaboration},
  {Price-Whelan}, {Sip{\H{o}}cz}, {G{\"u}nther}, {Lim}, {Crawford}, {Conseil},
  {Shupe}, {Craig}, {Dencheva}, {Ginsburg}, {VanderPlas}, {Bradley},
  {P{\'e}rez-Su{\'a}rez}, {de Val-Borro}, {Aldcroft}, {Cruz}, {Robitaille},
  {Tollerud}, {Ardelean}, {Babej}, {Bach}, {Bachetti}, {Bakanov}, {Bamford},
  {Barentsen}, {Barmby}, {Baumbach}, {Berry}, {Biscani}, {Boquien}, {Bostroem},
  {Bouma}, {Brammer}, {Bray}, {Breytenbach}, {Buddelmeijer}, {Burke},
  {Calderone}, {Cano Rodr{\'\i}guez}, {Cara}, {Cardoso}, {Cheedella}, {Copin},
  {Corrales}, {Crichton}, {D'Avella}, {Deil}, {Depagne}, {Dietrich}, {Donath},
  {Droettboom}, {Earl}, {Erben}, {Fabbro}, {Ferreira}, {Finethy}, {Fox},
  {Garrison}, {Gibbons}, {Goldstein}, {Gommers}, {Greco}, {Greenfield},
  {Groener}, {Grollier}, {Hagen}, {Hirst}, {Homeier}, {Horton}, {Hosseinzadeh},
  {Hu}, {Hunkeler}, {Ivezi{\'c}}, {Jain}, {Jenness}, {Kanarek}, {Kendrew},
  {Kern}, {Kerzendorf}, {Khvalko}, {King}, {Kirkby}, {Kulkarni}, {Kumar},
  {Lee}, {Lenz}, {Littlefair}, {Ma}, {Macleod}, {Mastropietro}, {McCully},
  {Montagnac}, {Morris}, {Mueller}, {Mumford}, {Muna}, {Murphy}, {Nelson},
  {Nguyen}, {Ninan}, {N{\"o}the}, {Ogaz}, {Oh}, {Parejko}, {Parley}, {Pascual},
  {Patil}, {Patil}, {Plunkett}, {Prochaska}, {Rastogi}, {Reddy Janga},
  {Sabater}, {Sakurikar}, {Seifert}, {Sherbert}, {Sherwood-Taylor}, {Shih},
  {Sick}, {Silbiger}, {Singanamalla}, {Singer}, {Sladen}, {Sooley},
  {Sornarajah}, {Streicher}, {Teuben}, {Thomas}, {Tremblay}, {Turner},
  {Terr{\'o}n}, {van Kerkwijk}, {de la Vega}, {Watkins}, {Weaver}, {Whitmore},
  {Woillez}, {Zabalza}, \& {Astropy Contributors}}]{10.3847/1538-3881/aabc4f}
{Astropy Collaboration}, {Price-Whelan}, A.~M., {Sip{\H{o}}cz}, B.~M., {et~al.}
  2018, \href{http://dx.doi.org/10.3847/1538-3881/aabc4f}{\color{magenta}\aj},
  \href{https://ui.adsabs.harvard.edu/abs/2018AJ....156..123A}{156, 123}

\bibitem[{{Astropy Collaboration} {et~al.}(2013){Astropy Collaboration},
  {Robitaille}, {Tollerud}, {Greenfield}, {Droettboom}, {Bray}, {Aldcroft},
  {Davis}, {Ginsburg}, {Price-Whelan}, {Kerzendorf}, {Conley}, {Crighton},
  {Barbary}, {Muna}, {Ferguson}, {Grollier}, {Parikh}, {Nair}, {Unther},
  {Deil}, {Woillez}, {Conseil}, {Kramer}, {Turner}, {Singer}, {Fox}, {Weaver},
  {Zabalza}, {Edwards}, {Azalee Bostroem}, {Burke}, {Casey}, {Crawford},
  {Dencheva}, {Ely}, {Jenness}, {Labrie}, {Lim}, {Pierfederici}, {Pontzen},
  {Ptak}, {Refsdal}, {Servillat}, \& {Streicher}}]{10.1051/0004-6361/201322068}
{Astropy Collaboration}, {Robitaille}, T.~P., {Tollerud}, E.~J., {et~al.} 2013,
  \href{http://dx.doi.org/10.1051/0004-6361/201322068}{\color{magenta}\aap},
  \href{https://ui.adsabs.harvard.edu/abs/2013A&A...558A..33A}{558, A33}

\bibitem[{{Aumann}(1985)}]{10.1086/131620}
{Aumann}, H.~H. 1985,
  \href{http://dx.doi.org/10.1086/131620}{\color{magenta}\pasp},
  \href{https://ui.adsabs.harvard.edu/abs/1985PASP...97..885A}{97, 885}

\bibitem[{{Aumann} {et~al.}(1984){Aumann}, {Gillett}, {Beichman}, {de Jong},
  {Houck}, {Low}, {Neugebauer}, {Walker}, \& {Wesselius}}]{10.1086/184214}
{Aumann}, H.~H., {Gillett}, F.~C., {Beichman}, C.~A., {et~al.} 1984,
  \href{http://dx.doi.org/10.1086/184214}{\color{magenta}\apjl},
  \href{https://ui.adsabs.harvard.edu/abs/1984ApJ...278L..23A}{278, L23}

\bibitem[{{Beuzit} {et~al.}(2019){Beuzit}, {Vigan}, {Mouillet}, {Dohlen},
  {Gratton}, {Boccaletti}, {Sauvage}, {Schmid}, {Langlois}, {Petit},
  {Baruffolo}, {Feldt}, {Milli}, {Wahhaj}, {Abe}, {Anselmi}, {Antichi},
  {Barette}, {Baudrand}, {Baudoz}, {Bazzon}, {Bernardi}, {Blanchard}, {Brast},
  {Bruno}, {Buey}, {Carbillet}, {Carle}, {Cascone}, {Chapron}, {Charton},
  {Chauvin}, {Claudi}, {Costille}, {De Caprio}, {de Boer}, {Delboulb{\'e}},
  {Desidera}, {Dominik}, {Downing}, {Dupuis}, {Fabron}, {Fantinel}, {Farisato},
  {Feautrier}, {Fedrigo}, {Fusco}, {Gigan}, {Ginski}, {Girard}, {Giro},
  {Gisler}, {Gluck}, {Gry}, {Henning}, {Hubin}, {Hugot}, {Incorvaia}, {Jaquet},
  {Kasper}, {Lagadec}, {Lagrange}, {Le Coroller}, {Le Mignant}, {Le Ruyet},
  {Lessio}, {Lizon}, {Llored}, {Lundin}, {Madec}, {Magnard}, {Marteaud},
  {Martinez}, {Maurel}, {M{\'e}nard}, {Mesa}, {M{\"o}ller-Nilsson}, {Moulin},
  {Moutou}, {Orign{\'e}}, {Parisot}, {Pavlov}, {Perret}, {Pragt}, {Puget},
  {Rabou}, {Ramos}, {Reess}, {Rigal}, {Rochat}, {Roelfsema}, {Rousset}, {Roux},
  {Saisse}, {Salasnich}, {Santambrogio}, {Scuderi}, {Segransan}, {Sevin},
  {Siebenmorgen}, {Soenke}, {Stadler}, {Suarez}, {Tiph{\`e}ne}, {Turatto},
  {Udry}, {Vakili}, {Waters}, {Weber}, {Wildi}, {Zins}, \&
  {Zurlo}}]{10.1051/0004-6361/201935251}
{Beuzit}, J.~L., {Vigan}, A., {Mouillet}, D., {et~al.} 2019,
  \href{http://dx.doi.org/10.1051/0004-6361/201935251}{\color{magenta}\aap},
  \href{https://ui.adsabs.harvard.edu/abs/2019A&A...631A.155B}{631, A155}

\bibitem[{{Binney} \& {Merrifield}(1998)}]{1998gaas.book.....B}
{Binney}, J. \& {Merrifield}, M. 1998, {Galactic Astronomy} (Princeton NJ:
  Princeton University Press)

\bibitem[{{Booth} {et~al.}(2013){Booth}, {Kennedy}, {Sibthorpe}, {Matthews},
  {Wyatt}, {Duch{\^e}ne}, {Kavelaars}, {Rodriguez}, {Greaves}, {Koning},
  {Vican}, {Rieke}, {Su}, {Moro-Mart{\'\i}n}, \&
  {Kalas}}]{10.1093/mnras/sts117}
{Booth}, M., {Kennedy}, G., {Sibthorpe}, B., {et~al.} 2013,
  \href{http://dx.doi.org/10.1093/mnras/sts117}{\color{magenta}\mnras},
  \href{https://ui.adsabs.harvard.edu/abs/2013MNRAS.428.1263B}{428, 1263}

\bibitem[{{Brunngr{\"a}ber} {et~al.}(2017){Brunngr{\"a}ber}, {Wolf},
  {Kirchschlager}, \& {Ertel}}]{10.1093/mnras/stw2675}
{Brunngr{\"a}ber}, R., {Wolf}, S., {Kirchschlager}, F., \& {Ertel}, S. 2017,
  \href{http://dx.doi.org/10.1093/mnras/stw2675}{\color{magenta}\mnras},
  \href{https://ui.adsabs.harvard.edu/abs/2017MNRAS.464.4383B}{464, 4383}

\bibitem[{{Burns} {et~al.}(1979){Burns}, {Lamy}, \&
  {Soter}}]{10.1016/0019-1035(79)90050-2}
{Burns}, J.~A., {Lamy}, P.~L., \& {Soter}, S. 1979,
  \href{http://dx.doi.org/10.1016/0019-1035(79)90050-2}{\color{magenta}\icarus},
  \href{https://ui.adsabs.harvard.edu/abs/1979Icar...40....1B}{40, 1}

\bibitem[{{Campo Bagatin} {et~al.}(1994){Campo Bagatin}, {Cellino}, {Davis},
  {Farinella}, \& {Paolicchi}}]{10.1016/0032-0633(94)90008-6}
{Campo Bagatin}, A., {Cellino}, A., {Davis}, D.~R., {Farinella}, P., \&
  {Paolicchi}, P. 1994,
  \href{http://dx.doi.org/10.1016/0032-0633(94)90008-6}{\color{magenta}\planss},
  \href{https://ui.adsabs.harvard.edu/abs/1994P&SS...42.1079C}{42, 1079}

\bibitem[{{Chen} {et~al.}(2008){Chen}, {Fitzgerald}, \&
  {Smith}}]{10.1086/592567}
{Chen}, C.~H., {Fitzgerald}, M.~P., \& {Smith}, P.~S. 2008,
  \href{http://dx.doi.org/10.1086/592567}{\color{magenta}\apj},
  \href{https://ui.adsabs.harvard.edu/abs/2008ApJ...689..539C}{689, 539}

\bibitem[{{Crifo} {et~al.}(1997){Crifo}, {Vidal-Madjar}, {Lallement}, {Ferlet},
  \& {Gerbaldi}}]{1997A&A...320L..29C}
{Crifo}, F., {Vidal-Madjar}, A., {Lallement}, R., {Ferlet}, R., \& {Gerbaldi},
  M. 1997, \aap,
  \href{https://ui.adsabs.harvard.edu/abs/1997A&A...320L..29C}{320, L29}

\bibitem[{{Curtis} {et~al.}(2005){Curtis}, {Rajaram}, {Toon}, \&
  {Tolbert}}]{10.1364/AO.44.004102}
{Curtis}, D.~B., {Rajaram}, B., {Toon}, O.~B., \& {Tolbert}, M.~A. 2005,
  \href{http://dx.doi.org/10.1364/AO.44.004102}{\color{magenta}\ao},
  \href{https://ui.adsabs.harvard.edu/abs/2005ApOpt..44.4102C}{44, 4102}

\bibitem[{{de Vries} {et~al.}(2012){de Vries}, {Acke}, {Blommaert}, {Waelkens},
  {Waters}, {Vandenbussche}, {Min}, {Olofsson}, {Dominik}, {Decin}, {Barlow},
  {Brandeker}, {di Francesco}, {Glauser}, {Greaves}, {Harvey}, {Holland},
  {Ivison}, {Liseau}, {Pantin}, {Pilbratt}, {Royer}, \&
  {Sibthorpe}}]{10.1038/nature11469}
{de Vries}, B.~L., {Acke}, B., {Blommaert}, J.~A.~D.~L., {et~al.} 2012,
  \href{http://dx.doi.org/10.1038/nature11469}{\color{magenta}\nat},
  \href{https://ui.adsabs.harvard.edu/abs/2012Natur.490...74D}{490, 74}

\bibitem[{{Dent} {et~al.}(2014){Dent}, {Wyatt}, {Roberge}, {Augereau},
  {Casassus}, {Corder}, {Greaves}, {de Gregorio-Monsalvo}, {Hales}, {Jackson},
  {Hughes}, {Lagrange}, {Matthews}, \& {Wilner}}]{10.1126/science.1248726}
{Dent}, W.~R.~F., {Wyatt}, M.~C., {Roberge}, A., {et~al.} 2014,
  \href{http://dx.doi.org/10.1126/science.1248726}{\color{magenta}Science},
  \href{https://ui.adsabs.harvard.edu/abs/2014Sci...343.1490D}{343, 1490}

\bibitem[{{Dohnanyi}(1969)}]{10.1029/JB074i010p02531}
{Dohnanyi}, J.~S. 1969,
  \href{http://dx.doi.org/10.1029/JB074i010p02531}{\color{magenta}\jgr},
  \href{https://ui.adsabs.harvard.edu/abs/1969JGR....74.2531D}{74, 2531}

\bibitem[{{Draine}(2003)}]{10.1086/379123}
{Draine}, B.~T. 2003,
  \href{http://dx.doi.org/10.1086/379123}{\color{magenta}\apj},
  \href{https://ui.adsabs.harvard.edu/abs/2003ApJ...598.1026D}{598, 1026}

\bibitem[{{Draine} \& {Lee}(1984)}]{10.1086/162480}
{Draine}, B.~T. \& {Lee}, H.~M. 1984,
  \href{http://dx.doi.org/10.1086/162480}{\color{magenta}\apj},
  \href{https://ui.adsabs.harvard.edu/abs/1984ApJ...285...89D}{285, 89}

\bibitem[{{Eiroa} {et~al.}(2010){Eiroa}, {Fedele}, {Maldonado},
  {Gonz{\'a}lez-Garc{\'\i}a}, {Rodmann}, {Heras}, {Pilbratt}, {Augereau},
  {Mora}, {Montesinos}, {Ardila}, {Bryden}, {Liseau}, {Stapelfeldt},
  {Launhardt}, {Solano}, {Bayo}, {Absil}, {Ar{\'e}valo}, {Barrado},
  {Beichmann}, {Danchi}, {Del Burgo}, {Ertel}, {Fridlund}, {Fukagawa},
  {Guti{\'e}rrez}, {Gr{\"u}n}, {Kamp}, {Krivov}, {Lebreton}, {L{\"o}hne},
  {Lorente}, {Marshall}, {Mart{\'\i}nez-Arn{\'a}iz}, {Meeus}, {Montes},
  {Morbidelli}, {M{\"u}ller}, {Mutschke}, {Nakagawa}, {Olofsson}, {Ribas},
  {Roberge}, {Sanz-Forcada}, {Th{\'e}bault}, {Walker}, {White}, \&
  {Wolf}}]{10.1051/0004-6361/201014594}
{Eiroa}, C., {Fedele}, D., {Maldonado}, J., {et~al.} 2010,
  \href{http://dx.doi.org/10.1051/0004-6361/201014594}{\color{magenta}\aap},
  \href{https://ui.adsabs.harvard.edu/abs/2010A&A...518L.131E}{518, L131}

\bibitem[{{Eiroa} {et~al.}(2013){Eiroa}, {Marshall}, {Mora}, {Montesinos},
  {Absil}, {Augereau}, {Bayo}, {Bryden}, {Danchi}, {del Burgo}, {Ertel},
  {Fridlund}, {Heras}, {Krivov}, {Launhardt}, {Liseau}, {L{\"o}hne},
  {Maldonado}, {Pilbratt}, {Roberge}, {Rodmann}, {Sanz-Forcada}, {Solano},
  {Stapelfeldt}, {Th{\'e}bault}, {Wolf}, {Ardila}, {Ar{\'e}valo}, {Beichmann},
  {Faramaz}, {Gonz{\'a}lez-Garc{\'\i}a}, {Guti{\'e}rrez}, {Lebreton},
  {Mart{\'\i}nez-Arn{\'a}iz}, {Meeus}, {Montes}, {Olofsson}, {Su}, {White},
  {Barrado}, {Fukagawa}, {Gr{\"u}n}, {Kamp}, {Lorente}, {Morbidelli},
  {M{\"u}ller}, {Mutschke}, {Nakagawa}, {Ribas}, \&
  {Walker}}]{10.1051/0004-6361/201321050}
{Eiroa}, C., {Marshall}, J.~P., {Mora}, A., {et~al.} 2013,
  \href{http://dx.doi.org/10.1051/0004-6361/201321050}{\color{magenta}\aap},
  \href{https://ui.adsabs.harvard.edu/abs/2013A&A...555A..11E}{555, A11}

\bibitem[{{Gaia Collaboration} {et~al.}(2021){Gaia Collaboration}, {Brown},
  {Vallenari}, {Prusti}, {de Bruijne}, {Babusiaux}, {Biermann}, {Creevey},
  {Evans}, {Eyer}, \& et~al.}]{10.1051/0004-6361/202039657}
{Gaia Collaboration}, {Brown}, A.~G.~A., {Vallenari}, A., {et~al.} 2021,
  \href{http://dx.doi.org/10.1051/0004-6361/202039657}{\color{magenta}\aap},
  \href{https://ui.adsabs.harvard.edu/abs/2021A&A...649A...1G}{649, A1}

\bibitem[{{Gaia Collaboration} {et~al.}(2016){Gaia Collaboration}, {Prusti},
  {de Bruijne}, {Brown}, {Vallenari}, {Babusiaux}, {Bailer-Jones}, {Bastian},
  {Biermann}, {Evans}, \& et~al.}]{10.1051/0004-6361/201629272}
{Gaia Collaboration}, {Prusti}, T., {de Bruijne}, J.~H.~J., {et~al.} 2016,
  \href{http://dx.doi.org/10.1051/0004-6361/201629272}{\color{magenta}\aap},
  \href{https://ui.adsabs.harvard.edu/abs/2016A&A...595A...1G}{595, A1}

\bibitem[{{Gardner} {et~al.}(2006){Gardner}, {Mather}, {Clampin}, {Doyon},
  {Greenhouse}, {Hammel}, {Hutchings}, {Jakobsen}, {Lilly}, {Long}, {Lunine},
  {McCaughrean}, {Mountain}, {Nella}, {Rieke}, {Rieke}, {Rix}, {Smith},
  {Sonneborn}, {Stiavelli}, {Stockman}, {Windhorst}, \&
  {Wright}}]{10.1007/s11214-006-8315-7}
{Gardner}, J.~P., {Mather}, J.~C., {Clampin}, M., {et~al.} 2006,
  \href{http://dx.doi.org/10.1007/s11214-006-8315-7}{\color{magenta}\ssr},
  \href{https://ui.adsabs.harvard.edu/abs/2006SSRv..123..485G}{123, 485}

\bibitem[{{Garnett}(1904)}]{10.1098/rsta.1904.0024}
{Garnett}, J.~C.~M. 1904,
  \href{http://dx.doi.org/10.1098/rsta.1904.0024}{\color{magenta}Philosophical
  Transactions of the Royal Society of London Series A},
  \href{https://ui.adsabs.harvard.edu/abs/1904RSPTA.203..385G}{203, 385}

\bibitem[{{Gray} {et~al.}(2006){Gray}, {Corbally}, {Garrison}, {McFadden},
  {Bubar}, {McGahee}, {O'Donoghue}, \& {Knox}}]{10.1086/504637}
{Gray}, R.~O., {Corbally}, C.~J., {Garrison}, R.~F., {et~al.} 2006,
  \href{http://dx.doi.org/10.1086/504637}{\color{magenta}\aj},
  \href{https://ui.adsabs.harvard.edu/abs/2006AJ....132..161G}{132, 161}

\bibitem[{{Greaves} {et~al.}(2005){Greaves}, {Holland}, {Wyatt}, {Dent},
  {Robson}, {Coulson}, {Jenness}, {Moriarty-Schieven}, {Davis}, {Butner},
  {Gear}, {Dominik}, \& {Walker}}]{10.1086/428348}
{Greaves}, J.~S., {Holland}, W.~S., {Wyatt}, M.~C., {et~al.} 2005,
  \href{http://dx.doi.org/10.1086/428348}{\color{magenta}\apjl},
  \href{https://ui.adsabs.harvard.edu/abs/2005ApJ...619L.187G}{619, L187}

\bibitem[{{Grigorieva} {et~al.}(2007){Grigorieva}, {Th{\'e}bault},
  {Artymowicz}, \& {Brandeker}}]{10.1051/0004-6361:20077686}
{Grigorieva}, A., {Th{\'e}bault}, P., {Artymowicz}, P., \& {Brandeker}, A.
  2007,
  \href{http://dx.doi.org/10.1051/0004-6361:20077686}{\color{magenta}\aap},
  \href{https://ui.adsabs.harvard.edu/abs/2007A&A...475..755G}{475, 755}

\bibitem[{{Harper} {et~al.}(2018){Harper}, {Runyan}, {Dowell}, {Wirth},
  {Amato}, {Ames}, {Amiri}, {Banks}, {Bartels}, {Benford}, {Berthoud},
  {Buchanan}, {Casey}, {Chapman}, {Chuss}, {Cook}, {Derro}, {Dotson}, {Evans},
  {Fixsen}, {Gatley}, {Guerra}, {Halpern}, {Hamilton}, {Hamlin}, {Hansen},
  {Heimsath}, {Hermida}, {Hilton}, {Hirsch}, {Hollister}, {Hostetter}, {Irwin},
  {Jhabvala}, {Jhabvala}, {Kastner}, {Kov{\'a}cs}, {Lin}, {Loewenstein},
  {Looney}, {Lopez-Rodriguez}, {Maher}, {Michail}, {Miller}, {Moseley},
  {Novak}, {Pernic}, {Rennick}, {Rhody}, {Sandberg}, {Sand ford}, {Santos},
  {Shafer}, {Sharp}, {Shirron}, {Siah}, {Silverberg}, {Sparr}, {Spotz},
  {Staguhn}, {Toorian}, {Towey}, {Tuttle}, {Vaillancourt}, {Voellmer},
  {Volpert}, {Wang}, \& {Wollack}}]{10.1142/S2251171718400081}
{Harper}, D.~A., {Runyan}, M.~C., {Dowell}, C.~D., {et~al.} 2018,
  \href{http://dx.doi.org/10.1142/S2251171718400081}{\color{magenta}Journal of
  Astronomical Instrumentation},
  \href{https://ui.adsabs.harvard.edu/abs/2018JAI.....740008H}{7, 1840008}

\bibitem[{{Harris} {et~al.}(2020){Harris}, {Millman}, {van der Walt},
  {Gommers}, {Virtanen}, {Cournapeau}, {Wieser}, {Taylor}, {Berg}, {Smith},
  {Kern}, {Picus}, {Hoyer}, {van Kerkwijk}, {Brett}, {Haldane}, {del R{\'\i}o},
  {Wiebe}, {Peterson}, {G{\'e}rard-Marchant}, {Sheppard}, {Reddy}, {Weckesser},
  {Abbasi}, {Gohlke}, \& {Oliphant}}]{10.1038/s41586-020-2649-2}
{Harris}, C.~R., {Millman}, K.~J., {van der Walt}, S.~J., {et~al.} 2020,
  \href{http://dx.doi.org/10.1038/s41586-020-2649-2}{\color{magenta}\nat},
  \href{https://ui.adsabs.harvard.edu/abs/2020Natur.585..357H}{585, 357}

\bibitem[{{H{\"a}{\ss}ner} {et~al.}(2018){H{\"a}{\ss}ner}, {Mutschke}, {Blum},
  {Zeller}, \& {Gundlach}}]{10.1093/mnras/sty2664}
{H{\"a}{\ss}ner}, D., {Mutschke}, H., {Blum}, J., {Zeller}, T., \& {Gundlach},
  B. 2018,
  \href{http://dx.doi.org/10.1093/mnras/sty2664}{\color{magenta}\mnras},
  \href{https://ui.adsabs.harvard.edu/abs/2018MNRAS.481.5022H}{481, 5022}

\bibitem[{{Holland} {et~al.}(2013){Holland}, {Bintley}, {Chapin},
  {Chrysostomou}, {Davis}, {Dempsey}, {Duncan}, {Fich}, {Friberg}, {Halpern},
  {Irwin}, {Jenness}, {Kelly}, {MacIntosh}, {Robson}, {Scott}, {Ade},
  {Atad-Ettedgui}, {Berry}, {Craig}, {Gao}, {Gibb}, {Hilton}, {Hollister},
  {Kycia}, {Lunney}, {McGregor}, {Montgomery}, {Parkes}, {Tilanus}, {Ullom},
  {Walther}, {Walton}, {Woodcraft}, {Amiri}, {Atkinson}, {Burger}, {Chuter},
  {Coulson}, {Doriese}, {Dunare}, {Economou}, {Niemack}, {Parsons},
  {Reintsema}, {Sibthorpe}, {Smail}, {Sudiwala}, \&
  {Thomas}}]{10.1093/mnras/sts612}
{Holland}, W.~S., {Bintley}, D., {Chapin}, E.~L., {et~al.} 2013,
  \href{http://dx.doi.org/10.1093/mnras/sts612}{\color{magenta}\mnras},
  \href{https://ui.adsabs.harvard.edu/abs/2013MNRAS.430.2513H}{430, 2513}

\bibitem[{{Hughes} {et~al.}(2018){Hughes}, {Duch{\^e}ne}, \&
  {Matthews}}]{10.1146/annurev-astro-081817-052035}
{Hughes}, A.~M., {Duch{\^e}ne}, G., \& {Matthews}, B.~C. 2018,
  \href{http://dx.doi.org/10.1146/annurev-astro-081817-052035}{\color{magenta}\araa},
  \href{https://ui.adsabs.harvard.edu/abs/2018ARA&A..56..541H}{56, 541}

\bibitem[{{Hunter}(2007)}]{10.1109/MCSE.2007.55}
{Hunter}, J.~D. 2007,
  \href{http://dx.doi.org/10.1109/MCSE.2007.55}{\color{magenta}Computing in
  Science and Engineering},
  \href{https://ui.adsabs.harvard.edu/abs/2007CSE.....9...90H}{9, 90}

\bibitem[{{Husser} {et~al.}(2013){Husser}, {Wende-von Berg}, {Dreizler},
  {Homeier}, {Reiners}, {Barman}, \&
  {Hauschildt}}]{10.1051/0004-6361/201219058}
{Husser}, T.~O., {Wende-von Berg}, S., {Dreizler}, S., {et~al.} 2013,
  \href{http://dx.doi.org/10.1051/0004-6361/201219058}{\color{magenta}\aap},
  \href{https://ui.adsabs.harvard.edu/abs/2013A&A...553A...6H}{553, A6}

\bibitem[{{Jovanovic} {et~al.}(2015){Jovanovic}, {Martinache}, {Guyon},
  {Clergeon}, {Singh}, {Kudo}, {Garrel}, {Newman}, {Doughty}, {Lozi}, {Males},
  {Minowa}, {Hayano}, {Takato}, {Morino}, {Kuhn}, {Serabyn}, {Norris},
  {Tuthill}, {Schworer}, {Stewart}, {Close}, {Huby}, {Perrin}, {Lacour},
  {Gauchet}, {Vievard}, {Murakami}, {Oshiyama}, {Baba}, {Matsuo}, {Nishikawa},
  {Tamura}, {Lai}, {Marchis}, {Duchene}, {Kotani}, \&
  {Woillez}}]{10.1086/682989}
{Jovanovic}, N., {Martinache}, F., {Guyon}, O., {et~al.} 2015,
  \href{http://dx.doi.org/10.1086/682989}{\color{magenta}\pasp},
  \href{https://ui.adsabs.harvard.edu/abs/2015PASP..127..890J}{127, 890}

\bibitem[{{Jura} {et~al.}(1998){Jura}, {Malkan}, {White}, {Telesco}, {Pina}, \&
  {Fisher}}]{10.1086/306184}
{Jura}, M., {Malkan}, M., {White}, R., {et~al.} 1998,
  \href{http://dx.doi.org/10.1086/306184}{\color{magenta}\apj},
  \href{https://ui.adsabs.harvard.edu/abs/1998ApJ...505..897J}{505, 897}

\bibitem[{{Kervella} {et~al.}(2004){Kervella}, {Th{\'e}venin}, {Morel},
  {Provost}, {Berthomieu}, {S{\'e} Gransan}, {Queloz}, {Bord{\'e}}, {di Folco},
  \& {Forveille}}]{2004IAUS..219...80K}
{Kervella}, P., {Th{\'e}venin}, F., {Morel}, P., {et~al.} 2004, in Stars as
  Suns : Activity, Evolution and Planets, ed. A.~K. {Dupree} \& A.~O. {Benz},
  Vol. 219 (Cambridge University Press),
  \href{https://ui.adsabs.harvard.edu/abs/2004IAUS..219...80K}{80}

\bibitem[{{Kim} {et~al.}(2018){Kim}, {Wolf}, {L{\"o}hne}, {Kirchschlager}, \&
  {Krivov}}]{10.1051/0004-6361/201833061}
{Kim}, M., {Wolf}, S., {L{\"o}hne}, T., {Kirchschlager}, F., \& {Krivov}, A.~V.
  2018,
  \href{http://dx.doi.org/10.1051/0004-6361/201833061}{\color{magenta}\aap},
  \href{https://ui.adsabs.harvard.edu/abs/2018A&A...618A..38K}{618, A38}

\bibitem[{{Kim} {et~al.}(2019){Kim}, {Wolf}, {Potapov}, {Mutschke}, \&
  {J{\"a}ger}}]{10.1051/0004-6361/201936014}
{Kim}, M., {Wolf}, S., {Potapov}, A., {Mutschke}, H., \& {J{\"a}ger}, C. 2019,
  \href{http://dx.doi.org/10.1051/0004-6361/201936014}{\color{magenta}\aap},
  \href{https://ui.adsabs.harvard.edu/abs/2019A&A...629A.141K}{629, A141}

\bibitem[{{Kirchschlager} \& {Wolf}(2013)}]{10.1051/0004-6361/201220486}
{Kirchschlager}, F. \& {Wolf}, S. 2013,
  \href{http://dx.doi.org/10.1051/0004-6361/201220486}{\color{magenta}\aap},
  \href{https://ui.adsabs.harvard.edu/abs/2013A&A...552A..54K}{552, A54}

\bibitem[{Kluyver {et~al.}(2016)Kluyver, Ragan-Kelley, P{\'e}rez, Granger,
  Bussonnier, Frederic, Kelley, Hamrick, Grout, Corlay, Ivanov, Avila, Abdalla,
  Willing, \& development team}]{jupyter}
Kluyver, T., Ragan-Kelley, B., P{\'e}rez, F., {et~al.} 2016, Jupyter Notebooks
  - a publishing format for reproducible computational workflows, ed.
  F.~{Loizides} \& B.~{Schmidt} (Netherlands: IOS Press), 87--90

\bibitem[{{Kobayashi} {et~al.}(2010){Kobayashi}, {Kimura}, {Yamamoto},
  {Watanabe}, \& {Yamamoto}}]{10.5047/eps.2009.03.001}
{Kobayashi}, H., {Kimura}, H., {Yamamoto}, S., {Watanabe}, S.~I., \&
  {Yamamoto}, T. 2010,
  \href{http://dx.doi.org/10.5047/eps.2009.03.001}{\color{magenta}Earth,
  Planets, and Space},
  \href{https://ui.adsabs.harvard.edu/abs/2010EP&S...62...57K}{62, 57}

\bibitem[{{Kobayashi} {et~al.}(2008){Kobayashi}, {Watanabe}, {Kimura}, \&
  {Yamamoto}}]{10.1016/j.icarus.2008.02.005}
{Kobayashi}, H., {Watanabe}, S.-i., {Kimura}, H., \& {Yamamoto}, T. 2008,
  \href{http://dx.doi.org/10.1016/j.icarus.2008.02.005}{\color{magenta}\icarus},
  \href{https://ui.adsabs.harvard.edu/abs/2008Icar..195..871K}{195, 871}

\bibitem[{{Krijt} \& {Kama}(2014)}]{10.1051/0004-6361/201423862}
{Krijt}, S. \& {Kama}, M. 2014,
  \href{http://dx.doi.org/10.1051/0004-6361/201423862}{\color{magenta}\aap},
  \href{https://ui.adsabs.harvard.edu/abs/2014A&A...566L...2K}{566, L2}

\bibitem[{{Krivov}(2010)}]{10.1088/1674-4527/10/5/001}
{Krivov}, A.~V. 2010,
  \href{http://dx.doi.org/10.1088/1674-4527/10/5/001}{\color{magenta}Research
  in Astronomy and Astrophysics},
  \href{https://ui.adsabs.harvard.edu/abs/2010RAA....10..383K}{10, 383}

\bibitem[{{Krivov} {et~al.}(2006){Krivov}, {L{\"o}hne}, \&
  {Srem{\v{c}}evi{\'c}}}]{10.1051/0004-6361:20064907}
{Krivov}, A.~V., {L{\"o}hne}, T., \& {Srem{\v{c}}evi{\'c}}, M. 2006,
  \href{http://dx.doi.org/10.1051/0004-6361:20064907}{\color{magenta}\aap},
  \href{https://ui.adsabs.harvard.edu/abs/2006A&A...455..509K}{455, 509}

\bibitem[{{Kurz} {et~al.}(2002){Kurz}, {Guilloteau}, \&
  {Shaver}}]{2002Msngr.107....7K}
{Kurz}, R., {Guilloteau}, S., \& {Shaver}, P. 2002, The Messenger,
  \href{https://ui.adsabs.harvard.edu/abs/2002Msngr.107....7K}{107, 7}

\bibitem[{{Lagage} {et~al.}(2004){Lagage}, {Pel}, {Authier}, {Belorgey},
  {Claret}, {Doucet}, {Dubreuil}, {Durand}, {Elswijk}, {Girardot}, {K{\"a}ufl},
  {Kroes}, {Lortholary}, {Lussignol}, {Marchesi}, {Pantin}, {Peletier},
  {Pirard}, {Pragt}, {Rio}, {Schoenmaker}, {Siebenmorgen}, {Silber}, {Smette},
  {Sterzik}, \& {Veyssiere}}]{2004Msngr.117...12L}
{Lagage}, P.~O., {Pel}, J.~W., {Authier}, M., {et~al.} 2004, The Messenger,
  \href{https://ui.adsabs.harvard.edu/abs/2004Msngr.117...12L}{117, 12}

\bibitem[{{Lebreton} {et~al.}(2012){Lebreton}, {Augereau}, {Thi}, {Roberge},
  {Donaldson}, {Schneider}, {Maddison}, {M{\'e}nard}, {Riviere-Marichalar},
  {Mathews}, {Kamp}, {Pinte}, {Dent}, {Barrado}, {Duch{\^e}ne}, {Gonzalez},
  {Grady}, {Meeus}, {Pantin}, {Williams}, \&
  {Woitke}}]{10.1051/0004-6361/201117714}
{Lebreton}, J., {Augereau}, J.~C., {Thi}, W.~F., {et~al.} 2012,
  \href{http://dx.doi.org/10.1051/0004-6361/201117714}{\color{magenta}\aap},
  \href{https://ui.adsabs.harvard.edu/abs/2012A&A...539A..17L}{539, A17}

\bibitem[{{Lecavelier Des Etangs} {et~al.}(1998){Lecavelier Des Etangs},
  {Vidal-Madjar}, \& {Ferlet}}]{1998A&A...339..477L}
{Lecavelier Des Etangs}, A., {Vidal-Madjar}, A., \& {Ferlet}, R. 1998, \aap,
  \href{https://ui.adsabs.harvard.edu/abs/1998A&A...339..477L}{339, 477}

\bibitem[{{Li} \& {Greenberg}(1998)}]{1998A&A...331..291L}
{Li}, A. \& {Greenberg}, J.~M. 1998, \aap,
  \href{https://ui.adsabs.harvard.edu/abs/1998A&A...331..291L}{331, 291}

\bibitem[{{Lisse} {et~al.}(2012){Lisse}, {Wyatt}, {Chen}, {Morlok}, {Watson},
  {Manoj}, {Sheehan}, {Currie}, {Thebault}, \&
  {Sitko}}]{10.1088/0004-637X/747/2/93}
{Lisse}, C.~M., {Wyatt}, M.~C., {Chen}, C.~H., {et~al.} 2012,
  \href{http://dx.doi.org/10.1088/0004-637X/747/2/93}{\color{magenta}\apj},
  \href{https://ui.adsabs.harvard.edu/abs/2012ApJ...747...93L}{747, 93}

\bibitem[{{L{\"o}hne}(2020)}]{10.1051/0004-6361/202037858}
{L{\"o}hne}, T. 2020,
  \href{http://dx.doi.org/10.1051/0004-6361/202037858}{\color{magenta}\aap},
  \href{https://ui.adsabs.harvard.edu/abs/2020A&A...641A..75L}{641, A75}

\bibitem[{{L{\"o}hne} {et~al.}(2012){L{\"o}hne}, {Augereau}, {Ertel},
  {Marshall}, {Eiroa}, {Mora}, {Absil}, {Stapelfeldt}, {Th{\'e}bault}, {Bayo},
  {Del Burgo}, {Danchi}, {Krivov}, {Lebreton}, {Letawe}, {Magain}, {Maldonado},
  {Montesinos}, {Pilbratt}, {White}, \& {Wolf}}]{10.1051/0004-6361/201117731}
{L{\"o}hne}, T., {Augereau}, J.~C., {Ertel}, S., {et~al.} 2012,
  \href{http://dx.doi.org/10.1051/0004-6361/201117731}{\color{magenta}\aap},
  \href{https://ui.adsabs.harvard.edu/abs/2012A&A...537A.110L}{537, A110}

\bibitem[{{MacGregor} {et~al.}(2017){MacGregor}, {Matr{\`a}}, {Kalas},
  {Wilner}, {Pan}, {Kennedy}, {Wyatt}, {Duchene}, {Hughes}, {Rieke}, {Clampin},
  {Fitzgerald}, {Graham}, {Holland}, {Pani{\'c}}, {Shannon}, \&
  {Su}}]{10.3847/1538-4357/aa71ae}
{MacGregor}, M.~A., {Matr{\`a}}, L., {Kalas}, P., {et~al.} 2017,
  \href{http://dx.doi.org/10.3847/1538-4357/aa71ae}{\color{magenta}\apj},
  \href{https://ui.adsabs.harvard.edu/abs/2017ApJ...842....8M}{842, 8}

\bibitem[{{MacGregor} {et~al.}(2013){MacGregor}, {Wilner}, {Rosenfeld},
  {Andrews}, {Matthews}, {Hughes}, {Booth}, {Chiang}, {Graham}, {Kalas},
  {Kennedy}, \& {Sibthorpe}}]{10.1088/2041-8205/762/2/L21}
{MacGregor}, M.~A., {Wilner}, D.~J., {Rosenfeld}, K.~A., {et~al.} 2013,
  \href{http://dx.doi.org/10.1088/2041-8205/762/2/L21}{\color{magenta}\apjl},
  \href{https://ui.adsabs.harvard.edu/abs/2013ApJ...762L..21M}{762, L21}

\bibitem[{{Macintosh} {et~al.}(2006){Macintosh}, {Graham}, {Palmer}, {Doyon},
  {Gavel}, {Larkin}, {Oppenheimer}, {Saddlemyer}, {Wallace}, {Bauman}, {Evans},
  {Erikson}, {Morzinski}, {Phillion}, {Poyneer}, {Sivaramakrishnan}, {Soummer},
  {Thibault}, \& {Veran}}]{10.1117/12.672430}
{Macintosh}, B., {Graham}, J., {Palmer}, D., {et~al.} 2006, in Society of
  Photo-Optical Instrumentation Engineers (SPIE) Conference Series, Vol. 6272,
  Society of Photo-Optical Instrumentation Engineers (SPIE) Conference Series,
  ed. B.~L. {Ellerbroek} \& D.~{Bonaccini Calia},
  \href{https://ui.adsabs.harvard.edu/abs/2006SPIE.6272E..0LM}{62720L}

\bibitem[{{Marino} {et~al.}(2018){Marino}, {Carpenter}, {Wyatt}, {Booth},
  {Casassus}, {Faramaz}, {Guzman}, {Hughes}, {Isella}, {Kennedy}, {Matr{\`a}},
  {Ricci}, \& {Corder}}]{10.1093/mnras/sty1790}
{Marino}, S., {Carpenter}, J., {Wyatt}, M.~C., {et~al.} 2018,
  \href{http://dx.doi.org/10.1093/mnras/sty1790}{\color{magenta}\mnras},
  \href{https://ui.adsabs.harvard.edu/abs/2018MNRAS.479.5423M}{479, 5423}

\bibitem[{{Matthews} {et~al.}(2014){Matthews}, {Krivov}, {Wyatt}, {Bryden}, \&
  {Eiroa}}]{10.2458/azu_uapress_9780816531240-ch023}
{Matthews}, B.~C., {Krivov}, A.~V., {Wyatt}, M.~C., {Bryden}, G., \& {Eiroa},
  C. 2014, in Protostars and Planets VI, ed. H.~{Beuther}, R.~S. {Klessen},
  C.~P. {Dullemond}, \& T.~{Henning},
  \href{https://ui.adsabs.harvard.edu/abs/2014prpl.conf..521M}{521}

\bibitem[{{Matthews} {et~al.}(2010){Matthews}, {Sibthorpe}, {Kennedy},
  {Phillips}, {Churcher}, {Duch{\^e}ne}, {Greaves}, {Lestrade}, {Moro-Martin},
  {Wyatt}, {Bastien}, {Biggs}, {Bouvier}, {Butner}, {Dent}, {di Francesco},
  {Eisl{\"o}ffel}, {Graham}, {Harvey}, {Hauschildt}, {Holland}, {Horner},
  {Ibar}, {Ivison}, {Johnstone}, {Kalas}, {Kavelaars}, {Rodriguez}, {Udry},
  {van der Werf}, {Wilner}, \& {Zuckerman}}]{10.1051/0004-6361/201014667}
{Matthews}, B.~C., {Sibthorpe}, B., {Kennedy}, G., {et~al.} 2010,
  \href{http://dx.doi.org/10.1051/0004-6361/201014667}{\color{magenta}\aap},
  \href{https://ui.adsabs.harvard.edu/abs/2010A&A...518L.135M}{518, L135}

\bibitem[{{Mie}(1908)}]{10.1002/andp.19083300302}
{Mie}, G. 1908,
  \href{http://dx.doi.org/10.1002/andp.19083300302}{\color{magenta}Annalen der
  Physik}, \href{https://ui.adsabs.harvard.edu/abs/1908AnP...330..377M}{330,
  377}

\bibitem[{{Minowa} {et~al.}(2010){Minowa}, {Hayano}, {Oya}, {Watanabe},
  {Hattori}, {Guyon}, {Egner}, {Saito}, {Ito}, {Takami}, {Garrel}, {Colley},
  {Golota}, \& {Iye}}]{10.1117/12.857818}
{Minowa}, Y., {Hayano}, Y., {Oya}, S., {et~al.} 2010, in Society of
  Photo-Optical Instrumentation Engineers (SPIE) Conference Series, Vol. 7736,
  Adaptive Optics Systems II, ed. B.~L. {Ellerbroek}, M.~{Hart}, N.~{Hubin}, \&
  P.~L. {Wizinowich},
  \href{https://ui.adsabs.harvard.edu/abs/2010SPIE.7736E..3NM}{77363N}

\bibitem[{{Morales} {et~al.}(2013){Morales}, {Bryden}, {Werner}, \&
  {Stapelfeldt}}]{10.1088/0004-637X/776/2/111}
{Morales}, F.~Y., {Bryden}, G., {Werner}, M.~W., \& {Stapelfeldt}, K.~R. 2013,
  \href{http://dx.doi.org/10.1088/0004-637X/776/2/111}{\color{magenta}\apj},
  \href{https://ui.adsabs.harvard.edu/abs/2013ApJ...776..111M}{776, 111}

\bibitem[{{Morales} {et~al.}(2016){Morales}, {Bryden}, {Werner}, \&
  {Stapelfeldt}}]{10.3847/0004-637X/831/1/97}
{Morales}, F.~Y., {Bryden}, G., {Werner}, M.~W., \& {Stapelfeldt}, K.~R. 2016,
  \href{http://dx.doi.org/10.3847/0004-637X/831/1/97}{\color{magenta}\apj},
  \href{https://ui.adsabs.harvard.edu/abs/2016ApJ...831...97M}{831, 97}

\bibitem[{{Moro-Martin} {et~al.}(2008){Moro-Martin}, {Wyatt}, {Malhotra}, \&
  {Trilling}}]{arXiv:astro-ph/0703383}
{Moro-Martin}, A., {Wyatt}, M.~C., {Malhotra}, R., \& {Trilling}, D.~E. 2008,
  in The Solar System Beyond Neptune, ed. M.~A. {Barucci}, H.~{Boehnhardt},
  D.~P. {Cruikshank}, \& A.~{Morbidelli} (Tucson: The University of Arizona
  Press),
  \href{https://ui.adsabs.harvard.edu/abs/2007astro.ph..3383M}{465--480}

\bibitem[{{Pawellek} \& {Krivov}(2015)}]{10.1093/mnras/stv2142}
{Pawellek}, N. \& {Krivov}, A.~V. 2015,
  \href{http://dx.doi.org/10.1093/mnras/stv2142}{\color{magenta}\mnras},
  \href{https://ui.adsabs.harvard.edu/abs/2015MNRAS.454.3207P}{454, 3207}

\bibitem[{{Pawellek} {et~al.}(2014){Pawellek}, {Krivov}, {Marshall},
  {Montesinos}, {{\'A}brah{\'a}m}, {Mo{\'o}r}, {Bryden}, \&
  {Eiroa}}]{10.1088/0004-637X/792/1/65}
{Pawellek}, N., {Krivov}, A.~V., {Marshall}, J.~P., {et~al.} 2014,
  \href{http://dx.doi.org/10.1088/0004-637X/792/1/65}{\color{magenta}\apj},
  \href{https://ui.adsabs.harvard.edu/abs/2014ApJ...792...65P}{792, 65}

\bibitem[{{Perez} \& {Granger}(2007)}]{10.1109/MCSE.2007.53}
{Perez}, F. \& {Granger}, B.~E. 2007,
  \href{http://dx.doi.org/10.1109/MCSE.2007.53}{\color{magenta}Computing in
  Science and Engineering},
  \href{https://ui.adsabs.harvard.edu/abs/2007CSE.....9c..21P}{9, 21}

\bibitem[{Petrenko \& Whitworth(2002)}]{physics_of_ice}
Petrenko, V. \& Whitworth, R. 2002, Physics of Ice (Oxford University Press)

\bibitem[{{Pilbratt} {et~al.}(2010){Pilbratt}, {Riedinger}, {Passvogel},
  {Crone}, {Doyle}, {Gageur}, {Heras}, {Jewell}, {Metcalfe}, {Ott}, \&
  {Schmidt}}]{10.1051/0004-6361/201014759}
{Pilbratt}, G.~L., {Riedinger}, J.~R., {Passvogel}, T., {et~al.} 2010,
  \href{http://dx.doi.org/10.1051/0004-6361/201014759}{\color{magenta}\aap},
  \href{https://ui.adsabs.harvard.edu/abs/2010A&A...518L...1P}{518, L1}

\bibitem[{{Poglitsch} {et~al.}(2010){Poglitsch}, {Waelkens}, {Geis},
  {Feuchtgruber}, {Vandenbussche}, {Rodriguez}, {Krause}, {Renotte}, {van
  Hoof}, {Saraceno}, {Cepa}, {Kerschbaum}, {Agn{\`e}se}, {Ali}, {Altieri},
  {Andreani}, {Augueres}, {Balog}, {Barl}, {Bauer}, {Belbachir}, {Benedettini},
  {Billot}, {Boulade}, {Bischof}, {Blommaert}, {Callut}, {Cara}, {Cerulli},
  {Cesarsky}, {Contursi}, {Creten}, {De Meester}, {Doublier}, {Doumayrou},
  {Duband }, {Exter}, {Genzel}, {Gillis}, {Gr{\"o}zinger}, {Henning},
  {Herreros}, {Huygen}, {Inguscio}, {Jakob}, {Jamar}, {Jean}, {de Jong},
  {Katterloher}, {Kiss}, {Klaas}, {Lemke}, {Lutz}, {Madden}, {Marquet},
  {Martignac}, {Mazy}, {Merken}, {Montfort}, {Morbidelli}, {M{\"u}ller},
  {Nielbock}, {Okumura}, {Orfei}, {Ottensamer}, {Pezzuto}, {Popesso},
  {Putzeys}, {Regibo}, {Reveret}, {Royer}, {Sauvage}, {Schreiber}, {Stegmaier},
  {Schmitt}, {Schubert}, {Sturm}, {Thiel}, {Tofani}, {Vavrek}, {Wetzstein},
  {Wieprecht}, \& {Wiezorrek}}]{10.1051/0004-6361/201014535}
{Poglitsch}, A., {Waelkens}, C., {Geis}, N., {et~al.} 2010,
  \href{http://dx.doi.org/10.1051/0004-6361/201014535}{\color{magenta}\aap},
  \href{https://ui.adsabs.harvard.edu/abs/2010A&A...518L...2P}{518, L2}

\bibitem[{{Pontoppidan} {et~al.}(2014){Pontoppidan}, {Salyk}, {Bergin},
  {Brittain}, {Marty}, {Mousis}, \&
  {{\"O}berg}}]{10.2458/azu_uapress_9780816531240-ch016}
{Pontoppidan}, K.~M., {Salyk}, C., {Bergin}, E.~A., {et~al.} 2014, in
  Protostars and Planets VI, ed. H.~{Beuther}, R.~S. {Klessen}, C.~P.
  {Dullemond}, \& T.~{Henning},
  \href{https://ui.adsabs.harvard.edu/abs/2014prpl.conf..363P}{363}

\bibitem[{{Potapov} {et~al.}(2018){Potapov}, {Mutschke}, {Seeber}, {Henning},
  \& {J{\"a}ger}}]{10.3847/1538-4357/aac6d3}
{Potapov}, A., {Mutschke}, H., {Seeber}, P., {Henning}, T., \& {J{\"a}ger}, C.
  2018, \href{http://dx.doi.org/10.3847/1538-4357/aac6d3}{\color{magenta}\apj},
  \href{https://ui.adsabs.harvard.edu/abs/2018ApJ...861...84P}{861, 84}

\bibitem[{{Poynting}(1904)}]{10.1098/rsta.1904.0012}
{Poynting}, J.~H. 1904,
  \href{http://dx.doi.org/10.1098/rsta.1904.0012}{\color{magenta}Philosophical
  Transactions of the Royal Society of London Series A},
  \href{https://ui.adsabs.harvard.edu/abs/1904RSPTA.202..525P}{202, 525}

\bibitem[{{Reinert} {et~al.}(2015){Reinert}, {Mutschke}, {Krivov}, {L{\"o}hne},
  \& {Mohr}}]{10.1051/0004-6361/201424276}
{Reinert}, C., {Mutschke}, H., {Krivov}, A.~V., {L{\"o}hne}, T., \& {Mohr}, P.
  2015,
  \href{http://dx.doi.org/10.1051/0004-6361/201424276}{\color{magenta}\aap},
  \href{https://ui.adsabs.harvard.edu/abs/2015A&A...573A..29R}{573, A29}

\bibitem[{{Remijan} {et~al.}(2020){Remijan}, {Biggs}, {Cortes}, {Dent}, {Di
  Francesco}, {Fomalont}, {Hales}, {Kameno}, {Mason}, {Philips}, {Saini},
  {Stoehr}, {Vila Vilaro}, \& {Villard}}]{alma_handbook_c8}
{Remijan}, A., {Biggs}, A., {Cortes}, P., {et~al.} 2020,
  \href{https://almascience.nrao.edu/documents-and-tools/cycle8/alma-technical-handbook}{ALMA
  Technical Handbook, ALMA Doc. 8.3, ver. 1.0}

\bibitem[{{Rieke} {et~al.}(2004){Rieke}, {Young}, {Engelbracht}, {Kelly},
  {Low}, {Haller}, {Beeman}, {Gordon}, {Stansberry}, {Misselt}, {Cadien},
  {Morrison}, {Rivlis}, {Latter}, {Noriega-Crespo}, {Padgett}, {Stapelfeldt},
  {Hines}, {Egami}, {Muzerolle}, {Alonso-Herrero}, {Blaylock}, {Dole}, {Hinz},
  {Le Floc'h}, {Papovich}, {P{\'e}rez-Gonz{\'a}lez}, {Smith}, {Su}, {Bennett},
  {Frayer}, {Henderson}, {Lu}, {Masci}, {Pesenson}, {Rebull}, {Rho}, {Keene},
  {Stolovy}, {Wachter}, {Wheaton}, {Werner}, \& {Richards}}]{10.1086/422717}
{Rieke}, G.~H., {Young}, E.~T., {Engelbracht}, C.~W., {et~al.} 2004,
  \href{http://dx.doi.org/10.1086/422717}{\color{magenta}\apjs},
  \href{https://ui.adsabs.harvard.edu/abs/2004ApJS..154...25R}{154, 25}

\bibitem[{{Robertson}(1937)}]{10.1093/mnras/97.6.423}
{Robertson}, H.~P. 1937,
  \href{http://dx.doi.org/10.1093/mnras/97.6.423}{\color{magenta}\mnras},
  \href{https://ui.adsabs.harvard.edu/abs/1937MNRAS..97..423R}{97, 423}

\bibitem[{{Rodigas} {et~al.}(2015){Rodigas}, {Stark}, {Weinberger}, {Debes},
  {Hinz}, {Close}, {Chen}, {Smith}, {Males}, {Skemer}, {Puglisi}, {Follette},
  {Morzinski}, {Wu}, {Briguglio}, {Esposito}, {Pinna}, {Riccardi}, {Schneider},
  \& {Xompero}}]{10.1088/0004-637X/798/2/96}
{Rodigas}, T.~J., {Stark}, C.~C., {Weinberger}, A., {et~al.} 2015,
  \href{http://dx.doi.org/10.1088/0004-637X/798/2/96}{\color{magenta}\apj},
  \href{https://ui.adsabs.harvard.edu/abs/2015ApJ...798...96R}{798, 96}

\bibitem[{{Strubbe} \& {Chiang}(2006)}]{10.1086/505736}
{Strubbe}, L.~E. \& {Chiang}, E.~I. 2006,
  \href{http://dx.doi.org/10.1086/505736}{\color{magenta}\apj},
  \href{https://ui.adsabs.harvard.edu/abs/2006ApJ...648..652S}{648, 652}

\bibitem[{{Temi} {et~al.}(2018){Temi}, {Hoffman}, {Ennico}, \&
  {Le}}]{10.1142/S2251171718400111}
{Temi}, P., {Hoffman}, D., {Ennico}, K., \& {Le}, J. 2018,
  \href{http://dx.doi.org/10.1142/S2251171718400111}{\color{magenta}Journal of
  Astronomical Instrumentation},
  \href{https://ui.adsabs.harvard.edu/abs/2018JAI.....740011T}{7, 1840011}

\bibitem[{{Thebault}(2016)}]{10.1051/0004-6361/201527626}
{Thebault}, P. 2016,
  \href{http://dx.doi.org/10.1051/0004-6361/201527626}{\color{magenta}\aap},
  \href{https://ui.adsabs.harvard.edu/abs/2016A&A...587A..88T}{587, A88}

\bibitem[{{Th{\'e}bault} \& {Augereau}(2007)}]{10.1051/0004-6361:20077709}
{Th{\'e}bault}, P. \& {Augereau}, J.~C. 2007,
  \href{http://dx.doi.org/10.1051/0004-6361:20077709}{\color{magenta}\aap},
  \href{https://ui.adsabs.harvard.edu/abs/2007A&A...472..169T}{472, 169}

\bibitem[{{Thebault} \& {Kral}(2019)}]{10.1051/0004-6361/201935341}
{Thebault}, P. \& {Kral}, Q. 2019,
  \href{http://dx.doi.org/10.1051/0004-6361/201935341}{\color{magenta}\aap},
  \href{https://ui.adsabs.harvard.edu/abs/2019A&A...626A..24T}{626, A24}

\bibitem[{{Th{\'e}bault} \& {Wu}(2008)}]{10.1051/0004-6361:20079133}
{Th{\'e}bault}, P. \& {Wu}, Y. 2008,
  \href{http://dx.doi.org/10.1051/0004-6361:20079133}{\color{magenta}\aap},
  \href{https://ui.adsabs.harvard.edu/abs/2008A&A...481..713T}{481, 713}

\bibitem[{{van Leeuwen}(2007)}]{10.1051/0004-6361:20078357}
{van Leeuwen}, F. 2007,
  \href{http://dx.doi.org/10.1051/0004-6361:20078357}{\color{magenta}\aap},
  \href{https://ui.adsabs.harvard.edu/abs/2007A&A...474..653V}{474, 653}

\bibitem[{{Warren}(1984)}]{10.1364/AO.23.001206}
{Warren}, S.~G. 1984,
  \href{http://dx.doi.org/10.1364/AO.23.001206}{\color{magenta}\ao},
  \href{https://ui.adsabs.harvard.edu/abs/1984ApOpt..23.1206W}{23, 1206}

\bibitem[{{Werner} {et~al.}(2004){Werner}, {Roellig}, {Low}, {Rieke}, {Rieke},
  {Hoffmann}, {Young}, {Houck}, {Brandl}, {Fazio}, {Hora}, {Gehrz}, {Helou},
  {Soifer}, {Stauffer}, {Keene}, {Eisenhardt}, {Gallagher}, {Gautier}, {Irace},
  {Lawrence}, {Simmons}, {Van Cleve}, {Jura}, {Wright}, \&
  {Cruikshank}}]{10.1086/422992}
{Werner}, M.~W., {Roellig}, T.~L., {Low}, F.~J., {et~al.} 2004,
  \href{http://dx.doi.org/10.1086/422992}{\color{magenta}\apjs},
  \href{https://ui.adsabs.harvard.edu/abs/2004ApJS..154....1W}{154, 1}

\bibitem[{{Wolf} \& {Voshchinnikov}(2004)}]{10.1016/j.cpc.2004.06.070}
{Wolf}, S. \& {Voshchinnikov}, N.~V. 2004,
  \href{http://dx.doi.org/10.1016/j.cpc.2004.06.070}{\color{magenta}Computer
  Physics Communications},
  \href{https://ui.adsabs.harvard.edu/abs/2004CoPhC.162..113W}{162, 113}

\bibitem[{{Wyatt}(2020)}]{10.1016/B978-0-12-816490-7.00016-3}
{Wyatt}, M. 2020, in The Trans-Neptunian Solar System, ed. D.~{Prialnik}, M.~A.
  {Barucci}, \& L.~{Young} (Elsevier),
  \href{https://ui.adsabs.harvard.edu/abs/2020tnss.book..351W}{351--376}

\bibitem[{{Wyatt} {et~al.}(2011){Wyatt}, {Clarke}, \&
  {Booth}}]{10.1007/s10569-011-9345-3}
{Wyatt}, M.~C., {Clarke}, C.~J., \& {Booth}, M. 2011,
  \href{http://dx.doi.org/10.1007/s10569-011-9345-3}{\color{magenta}Celestial
  Mechanics and Dynamical Astronomy},
  \href{https://ui.adsabs.harvard.edu/abs/2011CeMDA.111....1W}{111, 1}

\bibitem[{{Wyatt} \& {Whipple}(1950)}]{10.1086/145244}
{Wyatt}, S.~P. \& {Whipple}, F.~L. 1950,
  \href{http://dx.doi.org/10.1086/145244}{\color{magenta}\apj},
  \href{https://ui.adsabs.harvard.edu/abs/1950ApJ...111..134W}{111, 134}

\bibitem[{{Zook} \& {Berg}(1975)}]{10.1016/0032-0633(75)90078-1}
{Zook}, H.~A. \& {Berg}, O.~E. 1975, \planss,
  \href{https://ui.adsabs.harvard.edu/abs/1975P&SS...23..183Z}{23, 183}

\end{thebibliography}


\begin{appendix}

\section{Simulated spatially resolved images}\label{appdx_sim_images}

 To simulate the observation of a spatially resolved image, we convolve the
 image with a circular two-dimensional Gaussian to simulate the telescope beam.
 The beam size applied depends on the respective instrument and wavelength.
 For \textit{Herschel}/PACS we use values for the FWHM stated by
 \citet{10.1093/mnras/sts117} for DEBRIS survey images.
 For ALMA, the beam size varies with different telescope array configurations.
 We use values from the Cycle 8 Technical Handbook
 \citep[][Table 7.1]{alma_handbook_c8} for the C-1 configuration.
 By using a circular beam shape we assume the object to be observed at zenith.

 We extract radial profiles starting at the image center and sample them with
 multiples of the respective beam FWHM. This causes the number of data points
 of the radial profiles to vary with wavelength.
 See Table~\ref{table_beams_and_rp} for the adopted values of FWHM, the
 corresponding instrument and the number of radial profile data points.

 As an interferometric array, ALMA is only sensitive up to a maximum angular
 scale (named \textit{maximum recoverable scale} in the Technical Handbook)
 which depends on the smallest baseline in the array.
 Accurate modeling of ALMA observations is beyond the scope of
 this study. The ALMA wavelengths and beam sizes only serve as a general
 blueprint to model our synthetic observations. Thus, we neglect the maximum
 recoverable scale.
 \begin{table}
  \caption{Instrument beam sizes and number of radial profile data points}
  \label{table_beams_and_rp}
  \begin{tabular}{l l l l}
   \hline \hline
   \addlinespace[2pt]
   $\lambda \left[ \mu \textrm{m} \right]$ & FWHM [arcsec] & Instrument & Data points\\
    \hline
    \addlinespace[2pt]
    70 & 5.6 & \textit{Herschel}/PACS & 3\\
    160 & 11.4 & \textit{Herschel}/PACS & 2\\
    850 & 0.98 & ALMA/Band 7 & 13\\
    1300 & 1.47 & ALMA/Band 6 & 9\\
   \hline
  \end{tabular}
 \end{table}

\end{appendix}

\end{document}